\documentclass[aps,groupeaddress,prd,english,floatfix,preprint,nofootinbib]{revtex4-1}
\usepackage{epsfig}
\usepackage{amsmath}
\usepackage{slashed}
\usepackage{subfigure,graphicx}
\usepackage{array}

\newcommand{\beq}{\begin{equation}}
\newcommand{\eeq}{\end{equation}}
\newcommand{\beqa}{\begin{eqnarray}}
\newcommand{\eeqa}{\end{eqnarray}}
\newcommand{\bwt}{\begin{widetext}}
\newcommand{\ewt}{\end{widetext}}

\newcommand*{\ovl}{\overline}
\newcommand{\p}{\partial}

\newcommand*{\ra}{\rightarrow}

\newcommand*{\Lra}{\Longrightarrow}

\newcommand*{\eps}{\epsilon}

\newcommand*{\lam}{\lambda}

\newcommand{\nonr}{\nonumber}
\newcommand{\sigv}{\left< \sigma v\right>}

\begin{document}
\title{A Minimal Model of Majoronic Dark Radiation and  Dark Matter}

\author{We-Fu Chang}
\affiliation{Department of Physics, National Tsing Hua University,
HsinChu 300, Taiwan}
\author{John N. Ng}
\affiliation{Theory Group, TRIUMF, 4004 Wesbrook Mall, Vancouver BC V6T 2A3, Canada}

\date{\today}

\begin{abstract}
We extend the singlet Majoron model of dark radiation by adding another singlet scalar of unit lepton charge. The spontaneous breaking of global $U(1)_L$ connects
dark radiation with neutrino mass generation via type-I seesaw mechanism. The model naturally has a stable scalar dark matter field. It also predicts the
existence of a light scalar of mass less than 1 GeV that mixes with the Standard Model Higgs boson.  We perform a
numerical analysis of the parameters of the model by imposing constraints from giving correct relic abundance and satisfying bounds from direct dark matter detection, rare decays of B-meson, and invisible width of the Higgs boson. The viability of the model in accommodating the gamma rays from Galactic center is discussed as well. The model gives rise to new rare Higgs boson decays
such as four muons final states with displaced vertices. Another unique signal is two muons and missing energy recoil against the muon pair. Our result also
shows that such a bridge between dark radiation and the seesaw mechanism will put the seesaw scale in the range of 1-100 TeV.
\end{abstract}

\maketitle

\section{Introduction}
 The temperature fluctuation in the Cosmic Microwave Background Radiation (CMBR) is a sensitive measure of the number of relativistic degrees of freedom present
 before the era of recombination. This is usually given in terms of the effective number of neutrinos, $N_{\text{eff}}$, which in the Standard Model (SM) is three. Taking into account incomplete
 neutrino decoupling during $e^+ e^-$ annihilation and finite temperature effects leads to the SM prediction of $N_{\text{eff}}=3.046$ (see e.g. \cite{Pneff}). Observations thus far
 are consistent with this value.  However, recent measurements of CMBR from the Planck satellite~\cite{Planck} combined with that of the Hubble constant
from the Hubble Space Telescope~\cite{HST} resulted in a higher value of $N_{\text{eff}} = 3.83\pm 0.54$ at 95\%C.L. If one further includes data from WMAP9~\cite{W9}, Atacama Cosmology Telescope~\cite{ACT} and the South Pole Telescope~\cite{SPT} into the analysis, the extracted value becomes $N_{\text{eff}} = 3.62^{+0.50}_{-0.48}$ at 95\%C.L. The nonzero  $ \triangle N_{\text{eff}} \equiv N_{\text{eff}}-3.046$ can be taken as a hint of a dark radiation (DR) component beyond the expected three neutrino species at a confidence level of $2.4\sigma$. The origin and nature of this mysterious DR is not known. One possibility is a massless or nearly massless Goldstone boson arising from the spontaneous breaking of a $U(1)$ global symmetry. A Goldstone boson will count as 4/7 of a neutrino, and it appears to agree with observation. However, in order for the temperature of the Goldstone bosons to match with that of the neutrinos, they must remain in thermal equilibrium with ordinary matter
until muon annihilation\cite{Weinberg}. If Goldstone bosons decouple much earlier, they will contribute less than 4/7 to $N_{\text{eff}}$ as they will not be reheated but the neutrinos always will. Decoupling in the muon annihilation era yields a contribution $\triangle N_{\text{eff}} = 0.39$. It is definitely interesting to investigate the nature of this
global $U(1)$. Weinberg suggested that it is a new symmetry associated with the dark sector only. We believe it is worthwhile to investigate whether  this global $U(1)$  can be one of the
 well-known accidental symmetries of the SM, i.e. the baryon or the lepton number. In \cite{CNW14} we make use of $U(1)_L$, the global lepton number $L$, and its spontaneous breaking gives rise to the Goldstone boson which is the Majoron originally studied in \cite{Peccei}. This
allows us to make the connection between cosmic DR and neutrino mass generation
such as the seesaw mechanism \cite{seesaw}. In so doing we can  ask whether there are new constraints on the seesaw mechanism. Some other physics consequences are also studied in \cite{CNW14}. However, in this Majoronic DR model there is no dark matter (DM) candidate. In this paper we show that adding DM can be achieved while
maintaining much of the simplicity of the model.

In the Majoronic DR model a singlet Higgs field $S$ with lepton charge $L=2$ is utilized to give mass to the right-handed singlet neutrino $N_R$ by spontaneously
breaking $U(1)_L$. The imaginary or axial part of this scalar field is the
Goldstone boson which we identify as DR. In this paper we extend the model by adding a $L=1$ complex scalar field $\Phi$,  a genuine scalar field which does not
develop a vacuum expectation value (vev). After symmetry breaking a discrete $Z_2$
symmetry remains and we call that dark parity (DP). This parity will allow us to
identify the lightest of the two components of $\Phi$ as the DM candidate. In this case its stability is guaranteed by DP. The details of the model are given in the next section.

While Goldstone bosons are attractive candidates for DR, there are other possibilities studied in the literature. Light sterile neutrinos were considered in \cite{snu}. In addition, right-handed neutrinos with milliweak interactions as DR were attended to in \cite{RHN}. Contribution to $N_{\text{eff}}$ from axion-like particles was mentioned in \cite{NTY}. Connection of DR to asymmetric dark matter scenarios was studied in \cite{ADM}. A more unconventional view that $\Delta N_{\text{eff}}$ arises from not fully thermalized sub-eV light decay products of an exotic particle was studied in \cite{Hasenkamp:2014hma}.

This paper is organized as follows. In Sec.II we present the detailed construction of the model.
Sec.III is devoted to a calculation of the relic abundance of the DM particle and direct detection is discussed in Sec. IV.
This is followed by examination of the constraints on the parameter space of the model from direct detection, indirect detection and other experimental constraints.
The issue of  galactic diffuse gamma rays is taken up in Sec.VI. Since our model makes use of SM singlet scalars, it is not surprising that it will lead to new rare Higgs decays,
and this is studied in Sec.VII. Finally we give our conclusions in Sec.VIII.

\section{The Model}

We add to the particle contents of the SM a singlet Higgs field S which carries
lepton number $L=2$ and a non-Higgssed scalar field $\Phi$ with $L=1$. In order
to implement the type-I seesaw mechanism we add the requisite minimum of two singlet Majorana right-handed
neutrinos $N_i, i=1,2$. The new degrees of freedom together with the SM Higgs field $H$, lepton doublets $L_i, i=1,2,3$, and their quantum numbers are listed in Table I where
$L$ denotes the charge under a global $U(1)_{\text{L}}$ lepton symmetry.
\begin{center}
\begin{tabular}{|c|c|c|c|}
\hline
$\phantom{SU(2)}$&$\phantom{S}L\phantom{(2)}$&$SU(2)$ & $U(1)_Y$\\ \hline
$S$& 2\;\;& {$\mathbf {1}$} & {$\mathbf {0}$} \\ \hline
$\Phi$& 1\;\;& {$\mathbf {1}$}& {$\mathbf {0}$}\\\hline
H& 0\;\;& {$\mathbf{2}$} & {$\mathbf {\frac{1}{2}}$}\\ \hline
$N_{iR}$& 1\;\;& {$\mathbf {1}$} & {$\mathbf {0}$}\\ \hline
$L_i $& 1\;\;& {$\mathbf {2}$} & {$\mathbf {-\frac{1}{2}}$}  \\ \hline
\end{tabular}
\end{center}
\begin{center}
Table I. Relevant fields and their quantum numbers\end{center}.

With the quantum numbers assigned $\Phi$ will not have trilinear coupling with
$H$ and it will not contribute to the Majorana masses of $N_{Ri}$. It will not
have a Dirac mass type of couplings to the active neutrinos since it is a $SU(2)$ singlet. Thus, much of the Majoron model is not changed.

The scalar Lagrangian is
\begin{align}
\label{eq:lscalar}
{\mathcal {L}}_{scalar}&= (D_\mu H)^\dagger (D^\mu H) + (\p_\mu S)^\dagger (\p^\mu S) + (\p_\mu \Phi)^\dagger (\p^\mu \Phi) -V(H,S,\Phi)\,,  \nonr \\
V(H,S,\Phi)&= -\mu^2 H^\dagger H +\lambda (H^\dagger H)^2 -\mu_s^{2} S^\dagger S +\lambda_s (S^\dagger S)^2 +\lambda _{SH} (S^\dagger S)(H^\dagger H) \nonr \\
&{\phantom{=}} +m_{\Phi}^2 \Phi^\dagger \Phi + \lambda_{\Phi} (\Phi^\dagger \Phi)^2 +\lambda_{\Phi H}(\Phi^\dagger \Phi)(H^\dagger H) \nonr \\
&{\phantom{=}}+ \lambda_{\Phi S}(S^\dagger S)(\Phi^\dagger \Phi)+\frac{\kappa}{\sqrt 2}\left[ (\Phi^\dagger)^2 S + S^\dagger \Phi^2\right]\,,
\end{align}
and we take $\kappa$ to be real and $m_{\Phi}^2 >0$. Due to the $\kappa$ term it is more convenient to work with the usual linear representation of the
the scalar fields. We expand the fields as follow
\begin{align}
\Phi&= \frac{1}{\sqrt 2} (\rho + i\chi)\,, \nonr \\
S&= \frac{1}{\sqrt 2} (v_s +s + i\omega)\,,
\end{align}
and use the U-gauge for the Higgs field
\beq
H=\begin{pmatrix}0\\ \frac{v+h}{\sqrt 2}
\end{pmatrix}\,.
\eeq
The physical fields are $ \hat {S} =(h,s,\rho,\chi)$ and $\omega$ is the Goldstone boson which is the  Majoron. In the above basis the spin-0 mass matrix squared  is
\beq
\label{eq:smass}
\begin{split}
 M^2& = \\
& \begin{pmatrix} 2\lambda v^2 & \lambda_{SH}v v_s &0 &0 \\
\lambda_{SH} v v_s & 2\lambda_s v_s^{2} &0 &0 \\
0& 0& m^2_{\Phi} + \kappa v_s +\frac{1}{2}\lambda_{\Phi H}v^2+\frac{1}{2}\lambda_{\Phi S}v_{s}^2&0 \\
0 & 0 &0 & m_{\Phi}^2 - \kappa v_s+\frac{1}{2}\lambda_{\Phi H}v^2 +\frac{1}{2}\lambda_{\Phi S}v_{s}^2
\end{pmatrix}\,,
\end{split}
\eeq
and $\omega$ is massless. Note that the $\kappa$ term splits the masses of $\rho$ and $\chi$ and we require $m^2_{\Phi}>|\kappa v_s|-\frac{1}{2}(\lambda_{\Phi h}v^2+\lambda_{\Phi S}v_s^2)$.

The scalar potential becomes
\beq
\label{eq:scalarpot}
\begin{split}
V=& \frac{1}{2} \tilde{\hat{S}}M^2 \hat{S} +\lambda vh^3 +\frac{1}{4}\lambda h^4 + \lambda_s v_s s^3 +\lambda_s v_s \omega^2 s +\frac{1}{4}\lambda_s( s^4 +\omega^4)\\
&+ \frac{1}{2} \lambda_s \omega^2 s^2 +\frac{1}{2}\lambda_{SH} v_s s h^2 +\frac{1}{2}\lambda_{SH}v(s^2 +\omega^2)h +\frac{1}{4}\lambda_{SH}(s^2+\omega^2)h^2\\
&+\frac{1}{4}\lambda_{\Phi}(\rho^4 +\chi^4 +2\rho^2\chi^2)+ \frac{1}{2}\lambda_{\Phi H} v (\rho^2+\chi^2)h+ \frac{1}{4}\lambda_{\Phi H}(\rho^2 +\chi^2)h^2 \\
&+\frac{1}{2}\bar{\kappa} s\,\rho^2++\frac{1}{4}\lambda_{\Phi S}\left (s^2\rho^2+s^2\chi^2+\omega^2\rho^2+\omega^2\chi^2\right ) \\
&+\frac{1}{2}(\bar{\kappa} -2\kappa)s\,\chi^2  + \kappa \rho\chi \omega\,,
\end{split}
\eeq
where $\bar{\kappa}=\lambda_{\Phi S} v_s +\kappa$.
After spontaneous symmetry breaking of $U(1)_{\mathrm {L}}$ there remains a $Z_2$ symmetry which we refer to as DP. It is seen by the following transformation
\begin{align}
s,\omega,h &\longrightarrow s,\omega, h \nonr \\
\rho &\longrightarrow -\rho \nonr \\
\chi &\longrightarrow -\chi \,.
\end{align}
Our DP can be written as $(-1)^L$ which is coincidentally the same as the R-parity in supersymmetric models of DM.
Depending on the sign of $\kappa$, either $\rho$ or $\chi$ will be the dark matter candidate. For definiteness we choose $\kappa$ to be negative; thus
$\rho$ is our DM candidate. The field $\omega$ remains massless and is the Goldstone boson which will be the DR. The two remaining scalar bosons
are $s,h$. We can see from Eq.~\eqref{eq:smass} that they form a submatrix that can be diagonalized independently of $(\rho,\chi)$. They are analyzed in
Ref.~\cite{CNW07}, where the relevant Higgs bosons constraints were also presented. The mass squared eigenvalues are
\begin{equation}\label{eq:mevals}
m_{1,2}^2 = \lam v^2 + \lam_S v_S^2 \mp \sqrt{(\lam_S v_S^2 - \lam v^2)^2 + \lam_{HS}^2 v^2 v_S^2} \,.
\end{equation}
The physical mass eigenstates are then
\begin{equation}
\begin{pmatrix}
h_1 \\
h_2
\end{pmatrix}
=
\begin{pmatrix}
\cos\theta & -\sin\theta \\
\sin\theta & \cos\theta
\end{pmatrix}
\begin{pmatrix}
h \\
s
\end{pmatrix} \,,
\end{equation}
with mixing angle
\begin{equation}
\label{eq:thmix}
\tan 2\theta = \frac{\lam_{HS}v v_S}{\lam_S v_S^2 - \lam v^2}\, \,.
\end{equation}
We shall identify $h_1 \equiv h_{SM}$ as the SM Higgs, which was recently discovered at the LHC to have a mass of 125~GeV. Note that for small mixing (which shall be the case below), $m_{h_{SM}}^2 \approx 2\lam v^2$ and $m_2^2 \approx 2\lam_S v_S^2$. For all intent and purposes $h_1\approx h$ and $h_2\approx s$.

We can now employ the type-I seesaw mechanism to give masses to the active neutrinos. To set the notation we discuss the one family case which can be easily generalized to
the three families. The $U(1)_{\mathrm {L}}$ invariant interaction Lagrangian for the
neutrinos is
\beq
-{\mathcal {L}}_\ell= y\ovl{L}_L\tilde {H} N_R + Y \ovl{N_R^c} N_R S + h.c.\,,
\eeq
where $L= (n_L, e_L)^T$ is the SM lepton doublet and $\tilde {H}=i\sigma_2 H^*$. After symmetry breaking we get
\beq
-{\mathcal {L}}_\ell=\frac{y v}{\sqrt 2}\ovl{n_L}N_R + \frac{Y v_s}{\sqrt 2} \ovl{N^c_{R}} N_R + \frac{y}{\sqrt 2} \ovl{n_L} N_R h + \frac{Y}{\sqrt 2} (s+i\omega)\ovl{N^c_{R}}N_R +h.c.
\eeq
This yields the standard seesaw neutrino mass matrix
\beq
\begin{pmatrix} 0 & m\\m& M \end{pmatrix}\,,
\eeq
where $m=\frac{y v}{2\sqrt2} $ and $M =\frac{Y v_s}{\sqrt 2}$. For $\eps \equiv m_D/M \ll 1$, the standard type-I seesaw is operative. To leading order in $\eps$, the mass eigenstates are given by
\begin{equation}
\nu_L = n_L + \eps\,N_R^c \,, \qquad \eta_R = N_R - \eps\,n_L^c \,,
\end{equation}
with eigenvalues $m_\nu = \eps\,m_D$ and $M$, respectively, (after appropriate phase rotations). In order to obtain light active neutrino masses,
$m_\nu \lesssim 0.1$~eV, we require
\begin{equation}\label{eq:sscond}
y_1 = 2^{5/4}\left(\frac{m_\nu y_2 v_s}{v}\right)^{1/2} \lesssim 3.05 \times 10^{-6}\left(\frac{y_2 v_s}{\mathrm{TeV}}\right)^{1/2} \,.
\end{equation}
As a benchmark, we take $v_s = 1$~TeV and $y_2 = 1$. Then acceptable light neutrino masses can be obtained with $y_1$ the size of the electron Yukawa couplings, $y_e =\frac{\sqrt{2} m_e}{v} =2.91\times 10^{-6}$.

  Next we discuss how the neutrinos transform under DP. All SM leptons and  $N_R$ carry one lepton charge thus they are DP-odd.
  For the seesaw mechanism to operate,  $N_R$ needs to be heavy and will not be stable and thus cannot be a DM candidate.
Although the SM charged leptons will also be DP odd this does not lead to any new phenomenon since  the electroweak theory
is DP conserving, and  the electron remains stable.
Moreover, $\rho$ and $\chi$ do not have direct coupling  to the SM leptons.

 It is easy to check that the charged leptons do not couple to $\omega$ directly in the linear realization.
 In the nonlinear realization,  the $\omega$ couples to leptons derivatively and
  this interaction vanishes when one of the external leptons is an on-shell Dirac fermion.
In the linear representation  the process $f+\bar{f} \ra \omega\omega $ where $f$ is a charged lepton will proceed via the diagrams depicted
in Fig.{\ref{fig:fmaj}.
\begin{figure}[htbp]
\centering
\includegraphics[width=4.in]{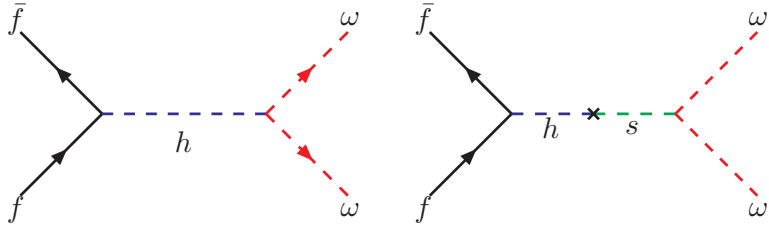}
\caption{Fermion-antifermion annihilation into a pair of Majorons.}
\label{fig:fmaj}
\end{figure}

Since we want the $\omega$ to act as the DR and gives $\Delta N_{\text {eff}}=.39$ it should decouple from the thermal bath around the muon annihilation  temperature\cite{Weinberg}. Earlier decoupling will not yield the above $\Delta N_{\text{eff}}$.
In that era, it is most convenient to calculate this using the mass insertion techniques.
We can take $s\simeq 4 m_f^2$ and  get
\beq
\label{eq:Lffmaj}
{\mathcal{L}}_{f\omega}\sim \frac{4 \lambda_{HS}m_f^3}{M_h^{2} M_s^{2}} \bar{f}f \omega\omega
\eeq
where $M_h^{2}\simeq 2\lambda v^2,\, M_s^2 \simeq 2\lambda_s\, v_s^{2}$,  and it agrees with the one obtained by using nonlinear realization \cite{Weinberg, CNW14} at low energies.  Equation\eqref{eq:Lffmaj} allows the $\omega$ to play the role of DR. For that it has to stay in thermal equilibrium until roughly the time of muon annihilations. This
requires the collision rate of $\omega$ into muons to be approximately the Hubble expansion rate at the
decoupling temperature $T_{dec}$,
\begin{equation}\label{eq:gb2DR}
\frac{\lam_{HS}^2 m_\mu^2 T_{dec}^5 m_{Pl}}{m_{h_{SM}}^4 m_{h_2}^4} \approx 1 \Lra m_{h_2} \approx 9.3\,\mathrm{GeV} \times (T_{dec}/m_\mu)^{5/4}\sqrt{|\lam_{HS}|} \,,
\end{equation}
where we take $m_{h_{SM}} = 125$~GeV. Hence we expect to have a $h_2$ much lighter then the Higgs which mixes with it. For notational simplicity  $h_1$ will be called $h$, and
$h_2$ will be called $s$.

Due to the $h-s$ mixing the Higgs boson acquires  three possible new two-body decays: (a) $h\ra \omega\omega$, (b) $ h\ra \rho\rho$ and (c) $h\ra s\, s$. Channel (b) will open if
$M_\rho < M_H/2$. (a) and (b) will add to the Higgs invisible width. As we shall see later we expect $M_s \ll M_H$ and whether (c) will lead to invisible decays depends on various parameters. Aside from those considerations  the widths for the above channels are
\begin{eqnarray}
\label{eq:Hwidths}
\Gamma(h\ra\omega\omega) &=& \frac{1}{32\pi}\frac{s_{\theta}^2 M_{h}^3}{v_s^2}\,, \nonr \\
\Gamma(h\ra\rho\rho)&=& \frac{1}{32\pi}\frac{M_\rho}{M_{h}^2}\sqrt{x_h-4}\left[\lambda_{\Phi H} v c_\theta -s_\theta\bar{\kappa}\right]^2\,, \nonr \\
\Gamma(h\ra s s)&=& \frac{1}{128\pi}\frac{M_\rho}{M_{h}^2}\sqrt{x_h-x_s}s_{2\theta}^2 \left(\frac{s_\theta}{v}-\frac{c_\theta}{v_s}\right)^2\left(M_{h}^2+2M_{s}^2\right)^2\,,
\end{eqnarray}
where $x_i=\frac{m_{i}^2}{M_\rho^2}$ and $i$ is the particle species. We also use the notation $s_\theta=\sin \theta$ and $c_\theta =\cos \theta$.

To get a qualitative feeling for the parameters we first take the case that only
(a) adds to the invisible Higgs width. From that the Higgs invisible decay
branching ratio is $\lesssim 0.19$ \cite{GKMRS13} with the Higgs width at about 4.1 MeV \cite{tome}, we get the Higgs invisible width to be $\lesssim 0.8 \ \mathrm {MeV}$. For small mixing, this yields the constraint
\beq
\lambda_{SH} < 0.0128\,.
\eeq
From Eq.\eqref{eq:gb2DR} we thus obtain $m_s\lesssim 1.05$ GeV.

If the $\rho$ channel is open we get instead
\beq
\label{eq:invh}
\frac{1}{2}\lambda_{\Phi H}^2 \sqrt{1-4/x_H}+\lambda_{SH}^2 < 1.27 \times 10^{-4}\,.
\eeq
This implies $\lambda_{\Phi H} \simeq \lambda_{SH}$. It is easy to see that scalar $s$ has mass of O(GeV) or less still holds qualitatively.

The signal from the decay $h\ra s s $ will depend on $M_s$, which dictates the decay modes of
$s$. The relevant modes are $s$ into light quarks and leptons, $\omega$'s and gluons. The invisible width is
\beq
\label{eq:sinv}
\Gamma(s\ra \omega\omega)=\frac{1}{32\pi}\frac{c_{\theta}^2 M_{s}^3}{v_s^2}\,,
\eeq
whereas the two fermions width is
\beq
\label{eq:s2f}
\Gamma(s\ra f\bar{f})=\frac{M_s}{8\pi}N_{c}^f\beta_{f}^3\left(\frac{m_f s_\theta}{v}\right)^2\,,
\eeq
where $\beta_f =\sqrt{ 1-\frac{4m_f^2}{M_s^2}}$ and $N_{c}^f$ denotes the color of the fermion.

How large a contribution of this to the Higgs invisible decay depends on the relative size of $\frac{M_s}{v_s}$ and $\theta$.
Nevertheless it is clear this will not change the result $M_s \lesssim {\mathrm {O (GeV)}}$. For $M_s\lesssim 1 $ GeV we also have
\beq
\label{eq:s_decay_ratio}
\Gamma_{\mu^+\mu^-}:\Gamma_{u\bar{u},d\bar{d}}:\Gamma_{gg}=m_\mu^2\beta_\mu^3:3m^{2}_{u,d}\beta_{\pi}^2 : \left(\frac{\alpha_s}{\pi}\right)^2 M_s^2\left(\frac{6-2\beta_{\pi}^3}{3}\right)^2\,,
\eeq
where we have neglected the kaon modes which are kinematically suppressed.
To close this section we mention that some  low-energy consequences of this light scalar have been explored in \cite{CNW14}.

\section{ Dark Matter and Its Relic Abundance}
\subsection{ DM annihilation channels}
In our model, due to the $Z_2$ DP the lighter of $\rho$ and $\chi$ will be the DM.
Without loss of generality we choose it to be $\rho$. Then $\chi$ can decay into
$\rho$ and $\omega$. Hence there is only one DM candidate. Note that $M^2_{\rho}-M^2_{\chi}=2\kappa v_s$ and the mass difference is not necessarily small. The relic density of $\rho$ can be calculated by evaluating the rate
of a pair of $\rho$ annihilating into SM particles as well as new scalars that are lighter than $\rho$. The SM channels are depicted in Fig.(\ref{fig:DMtoSM})
\begin{figure}[htbp]
\centering
\includegraphics[width=4.in]{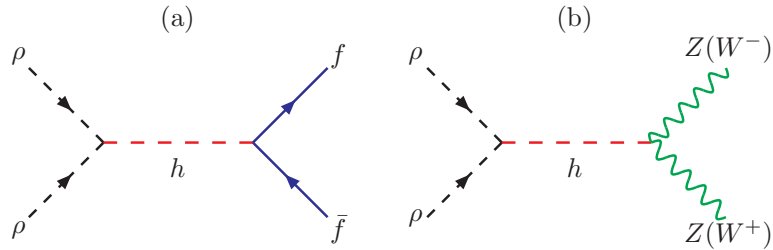}
\caption{$\rho\rho$ annihilation into SM particles}
\label{fig:DMtoSM}
\end{figure}
 and are open if it is heavy enough. Two DM's can annihilate into lighter scalars as well as the Majorons. These reactions are  given below.
\begin{figure}[htbp]
\centering
\includegraphics[width=4.in]{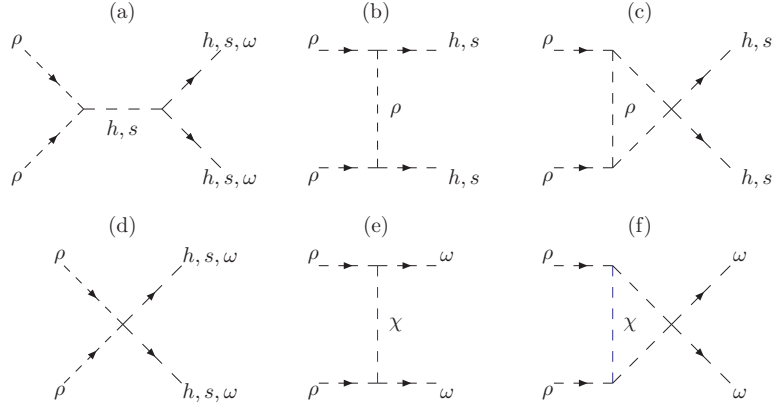}
\caption{$\rho\rho$ to a pair of SM Higgs, light scalars and Majorons}
\label{fig:DMtoscalars}
\end{figure}
Since the mixing between the Higgs and the light scalar is small, we can neglect it here and only the diagonal terms are important. We note that
there can also be the coannihilation of $\rho$ and $\chi$ into scalars and Majoron but these will require $\kappa$ to be fine-tuned to very small values.
The effect of the neutrino sector on DM relic abundance depends on the mass $M_R$ of $N_R$. We are interested in the case of $M_\rho< M_R$ then the neutrino sector has minimal effect on DM relic abundance.
\subsection{Relic Density}
The evolution of the comoving particle density is given by the Boltzmann equation
\begin{equation}
\frac{1}{n_{eq}}\frac{\p{n}}{\p t}=\Gamma\cdot\left(\frac{n^{2}}{n^{2}_{eq}}-1\right)-3H\frac{n}{n_{eq}}\,,
\end{equation}
where $n$ is the particle density at time $t$ and $n_{eq}$ is the density at equilibrium, $H$ is Hubble expansion rate and $\Gamma$ parametrizes the interaction rate, $\Gamma=\left<\sigma v\right>n_{eq}$ with $\left<\sigma v\right>$ the thermally averaged annihilation cross section. By solving numerically the above equation one can find the temperature at which particles depart from equilibrium and freeze out. Crudely
speaking since time is inversely proportional to temperature the above equation can be viewed as  an evolution equation with respect to temperature. The freeze-out temperature $T_f$ is given by
\beq
\label{eq:xFO}
x_f\equiv \frac{M_\rho}{T_f}=\ln\left(0.038 g_X \left<\sigma v\right> M_\rho\,M_{\text{Pl}}\sqrt{\frac{x_f}{g_*}}\right)\,,
\eeq
where $M_{\text{Pl}}$ is the Planck mass and $g_*$ is the effective number of relativistic degrees of freedom at temperature $T$.
For large $x_f \sim 20$ one can neglect the $x_f$ factor in the logarithm. Once we know $\left<\sigma v\right>$, we can
calculate the freeze-out temperature of $X$ with a given mass.

It is now straightforward to calculate the $\rho \rho$ annihilation cross sections to various final states. The Feynman diagrams are given in Figs.(\ref{fig:DMtoSM},\ref{fig:DMtoscalars}).
For completeness, the results we get for a general mixing  are
\begin{eqnarray}
\label{eq:approsigv}
( \sigma v )_{ss} &=& \frac{1}{64\pi}\frac{\sqrt{1-x_s}}{M_\rho ^2}\left[ q_{SS} +{g_{\rho H} \lambda_{HSS}\over M_\rho^2(4-x_H) }
+{g_{\rho S} \lambda_{SSS}\over M_\rho^2(4-x_S) }-\frac{2g_{\rho S}^2}{M_{\rho}^2(2-x_s)}\right]^2 \,, \nonr \\
( \sigma v )_{HH} &=& \frac{1}{64\pi}\frac{\sqrt{1-x_H }}{M_\rho ^2}\left[ q_{HH}+  \frac{g_{\rho H} \lambda_{HHH}}{M_\rho^2(4-x_H)}+ \frac{g_{\rho S} \lambda_{SHH}}{M_\rho^2(4-x_S)}-\frac{2 g_{\rho H}^2 }{M_\rho^2 (2-x_H)}\right]^2\,, \nonr \\
( \sigma v )_{Hs} &=&\frac{1}{32\pi}\frac{\Delta}{M_{\rho}^2}\left[ q_{HS}+ \frac{g_{\rho H} \lambda_{SHH} }{M_{\rho}^2(4-x_H)} + \frac{g_{\rho S} \lambda_{HSS} }{M_{\rho}^2(4-x_S)}
-\frac{4 g_{\rho H} g_{\rho S}}{M_{\rho}^2(4-x_H -x_s)}\right]^2\,, \nonr \\
( \sigma v )_{\omega\omega}&=& \frac{1}{64\pi M_{\rho}^2}\left[\frac{g_{\rho H} g_{\omega H}}{M_\rho^2 (4-x_H)}+\frac{g_{\rho S} g_{\omega S}}{M_\rho^2 (4-x_S)}+\lambda_{\Phi S}-\frac{2\kappa ^2}{M_\rho ^2(1+x_\chi)}\right]^2\,, \nonr \\
( \sigma v )_{WW} &=&\frac{1}{8\pi}\frac{\lambda_{\Phi H}^2}{M_\rho ^2}\sqrt{1-x_W}\left[4-4x_W +3x_{W}^2\right] \left[\frac{c_\theta^2}{(4-x_H)} +\frac{s_\theta^2}{(4-x_S)} \right]^2 \,,\nonr \\
( \sigma v )_{ZZ} &=& \frac{1}{16\pi}\frac{\lambda_{\Phi H}^2}{M_\rho ^2}\sqrt{1-x_Z}\left[4-4x_Z +3x_{Z}^2\right] \left[\frac{c_\theta^2}{(4-x_H)} +\frac{s_\theta^2}{(4-x_S)} \right]^2 \,,\nonr \\
( \sigma v )_{f\bar{f}}&=& \frac{N_c}{4\pi}\frac{\lambda_{\Phi H}^2 x_f}{M_\rho ^2}(1-x_f)^{\frac{3}{2}}
\left[\frac{c_\theta^2}{(4-x_H)} +\frac{s_\theta^2}{(4-x_S)} \right]^2\,,
\end{eqnarray}
where $\Delta^2=1+\frac{1}{16}x^2_{H}+\frac{1}{16}x^2_{s}-\frac{1}{8}x_H x_s -\frac{1}{2}x_H-\frac{1}{2}x_s$, $x_i=\frac{M_i ^2}{M_\rho ^2}$ for $i=W,Z,H,f,S,\chi$, and the subscripts
denote the final state. The coupling in the scalar mass basis are given as
\beqa
q_{SS}&=&\lambda_{\Phi S} c^2_\theta +\lambda_{\Phi H} s^2_\theta\,,\;
q_{HH}=\lambda_{\Phi S} s^2_\theta +\lambda_{\Phi H} c^2_\theta\,,\;
q_{HS}=(\lambda_{\Phi H}-\lambda_{\Phi S}) c_\theta s_\theta\,,\nonr\\
g_{\rho S}&=& \bar{\kappa} c_\theta +\lambda_{\Phi H} v s_\theta\,,\;
g_{\rho H}= -\bar{\kappa} s_\theta +\lambda_{\Phi H} v c_\theta\,,\nonr\\
g_{\omega S}&=& \lambda_{S H} v c_\theta -2 \lambda_{S} v_S s_\theta\,,\;
g_{\omega H}= \lambda_{S H} v s_\theta + 2 \lambda_{S} v_S c_\theta\,,\nonr\\
\lambda_{HHH}&=& 6\lambda v c_\theta^3 -6\lambda_S v_S s_\theta^3 + 3\lambda_{SH}  s_\theta c_\theta (v s_\theta -v_S c_\theta)\,,\nonr\\
\lambda_{SSS}&=& 6\lambda v s_\theta^3 +6\lambda_S v_S c_\theta^3 + 3\lambda_{SH}  s_\theta c_\theta (v_S s_\theta + v c_\theta)\,,\nonr\\
\lambda_{SHH}&=& 6 s_\theta c_\theta( \lambda v c_\theta +\lambda_S v_S s_\theta ) + \lambda_{SH} v_S( c_\theta^3-2 s_\theta^2 c_\theta)
+ \lambda_{SH} v ( s_\theta^3-2 s_\theta c_\theta^2)\,,\nonr\\
\lambda_{HSS}&=& 6 s_\theta c_\theta( \lambda v s_\theta -\lambda_S v_S c_\theta ) + \lambda_{SH} v_S(-s_\theta^3+2 s_\theta c_\theta^2)
+ \lambda_{SH} v ( c_\theta^3-2 s_\theta^2 c_\theta)\,.
\eeqa
 For high temperatures these will give $\sigv$. It is well known that in order to get the
correct relic density the total $\sigv$ should be approximately $3\times 10^{-26}
{\mathrm{cm}}^3/\mathrm{s}$. Due to the number of unknown parameters a numerical scan is required for the correct relic density.
This will be given in Sec. V.

\section{ Direct Detection}
The DM candidate could be detected by measuring the energy deposited in a low background detector by the scattering of $\rho$ with a nucleus of the detector. Since $\rho$ is a scalar there are only spin independent scattering via t-channel exchange of virtual $h$ and $s$. This is depicted in Fig.(\ref{fig:DMn}).
\begin{figure}[htbp]
\centering
\includegraphics[width=2.in]{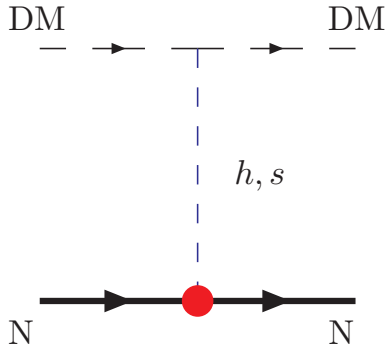}
\caption{Leading channel for DM-nucleon scattering via Higgs and light scalar exchange. DM is $\rho$}
\label{fig:DMn}
\end{figure}
The cross section  is
\beq
\label{eq:DMnucleon}
\sigma_{\rho \mathrm{n}}=\frac{G_F M_{n}^2 \eta^2 m_r^2(n,\rho)}{4\sqrt{2}\pi M_{\rho}^2 M_{H}^2 \lambda}\left[\lambda_{\Phi H}\left(c^2_{\theta}+s^2_{\theta}\left(\frac{M_h}{M_s}\right)^2 \right)-s_\theta c_\theta\frac{\bar{\kappa}}{v}\left(1-\left(\frac{M_h}{M_s}\right)^2\right)\right]^2\,,
\eeq
where the reduced mass is
\beq
m_r(n,\rho)=\frac{M_\rho M_n}{M_\rho + M_n}\,,
\eeq
and $M_n$ is the nucleon mass. For a qualitative estimation we take  $\eta=0.3$ in our numerical analysis and
ignore all the possible effects from isospin breaking or the strange-quark content which can be accounted for( see for example\cite{Crivellin:2013ipa} ). Since $s$ is very light compared to the Higgs boson its contribution cannot be neglected. Hence, the direct detection sets a strong constraint on the parameters combination $\frac{\bar{\kappa}}{v}s_{2\theta}$.

\section{Parameters Scan and Numerical Analysis}
The scalar potential introduces 8 more parameters to the SM.  We perform a global numerical  scan to investigate the general properties
of this model in different regions of parameter space. We employ 4000 randomly generated points.
The parameters scan is performed for $m_\rho \in [6, 2000]$ GeV as follow:
\begin{enumerate}
\item  The mass of light scalar $M_s$ is randomly chosen between $0.0$ and $1.0$ GeV. Such a light scalar is required if the Goldstone is associated with a dark $U(1)$ global symmetry.
\item  So as  not to miss any possible solution, the mixing, $\sin\theta$, is randomly picked between  $\pm 0.01$. This value is dictated by constraints on light scalars mixing with the Higgs from rare B-meson decays \cite{BKs}.
With the above inputs, we fix $|\lambda_{SH}| = (M_s/22.11 \mbox{GeV})^2$ by the requirement that the Majoron decouples from the primordial plasma at around twice the muon mass ($T_{dec}\sim 2 m_\mu$). This does not change $\Delta N_{\text{eff}}=.39$ as compared to using $T_{dec}\sim m_\mu$ \cite{Weinberg} and allows us to probe a larger parameter space. Also its sign is opposite to that of $\sin\theta$. This is to be viewed as a benchmark point and its exact value is  unknown since it depends on the actual decoupling temperature.
From the mass diagonalization, two parameters in the scalar potential and $v_S$ can be expressed in terms of mass eigenvalues, $M_H=125$ GeV, $M_s$, and the mixing:
\beq
\lambda=  {(M_H^2 c_\theta^2 +M_s^2 s^2_\theta)\over 2v^2  }\,,\;
\lambda_S=  {(M_s^2 c^2_\theta +M_H^2 s^2_\theta)\over 2v_S^2  }\,,\;
v_S=\frac{s_\theta c_\theta}{\lambda_{SH} v }(M_s^2-M_H^2).
\eeq
\item Next, we allow $\bar{\kappa}$ to be randomly chosen in the region between $- v$ and $+v$.
\item Then,  $\lambda_{\Phi S}$ is randomly chosen between $-4\sqrt{\pi \lambda_S}$ and $ 4\pi $.
Since we limit our discussion to the perturbative  regime so the upper bound of  any dimensionless coupling is set to be $4\pi$.
The lower bound is derived from that $(4 \lambda_S \lambda_\Phi - \lambda_{\Phi S}^2)>0$, which is the positivity  requirement of the scalar potential, with  the largest $ \lambda_\phi =4\pi$.
And it is further required to satisfy  the condition that $\kappa = \bar{\kappa}-\lambda_{\phi S} v_s <0$.
That $\kappa$ is negative  is because we pick  $\rho$ to be the dark matter.
And the mass of $\chi$ is determined to be $M_\chi=\sqrt{M_\rho^2-2 \kappa v_S}$.
\item  Finally, we allow $\lambda_{\phi H}$ to be randomly chosen between $-4\sqrt{\pi \lambda}$ and $4\pi$ for the same reason as in the case of   $\lambda_{\phi S}$.
\item$\lambda_{\Phi}$ does not enter into the calculations of the observables here. It remains unconstrained .
\end{enumerate}
The program will register the points which satisfy all the following four criteria:
\begin{itemize}
\item $(M_\rho^2 +M_\chi^2 -\lambda_{\Phi H}^2 v^2 -\lambda_{\Phi S}^2 v_S^2) >0$ so that $ M_\Phi^2 >0$.
\item The SM Higgs invisible decay width $\Gamma^h_{inv}<0.8$ MeV.
\item The thermal average annihilation cross section is within the range $(2.5\pm 0.1)\times 10^{-9} (\mbox{GeV})^{-2}$.
\item The  spin-independent elastic $\rho$-nucleon scattering cross section, Eq.(\ref{eq:DMnucleon}), is smaller than the LUX $90\%$ confidence limit \cite{LUX}.
\end{itemize}

First of all, we found that it is less probable to find  solutions
with very small mixing angle. Moreover, even we allow the $M_S$ to be chosen between  $0.0$ and $1.0$ GeV,
the resulting $M_S$ is cut off at around $0.8$ GeV with a smooth distribution peaks at around $0.4$ GeV.  Both $\theta$ and $M_S$  are insensitive to $M_\rho$, see Fig.\ref{fig:thetaMs_distribution}.
  \begin{figure}
\begin{center}
\includegraphics[width=0.46\textwidth]{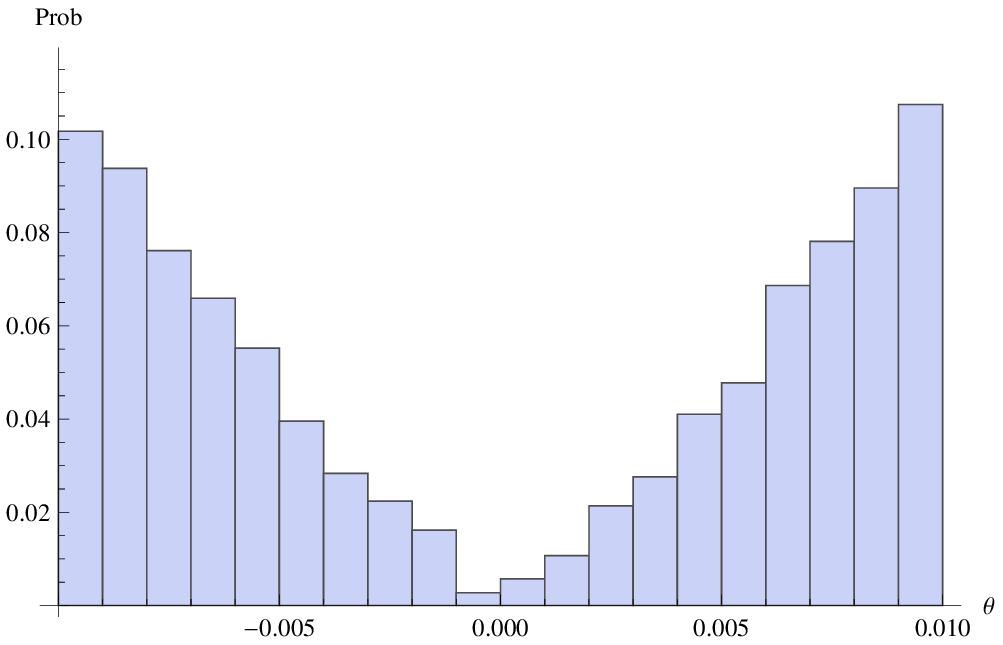}
\includegraphics[width=0.46\textwidth]{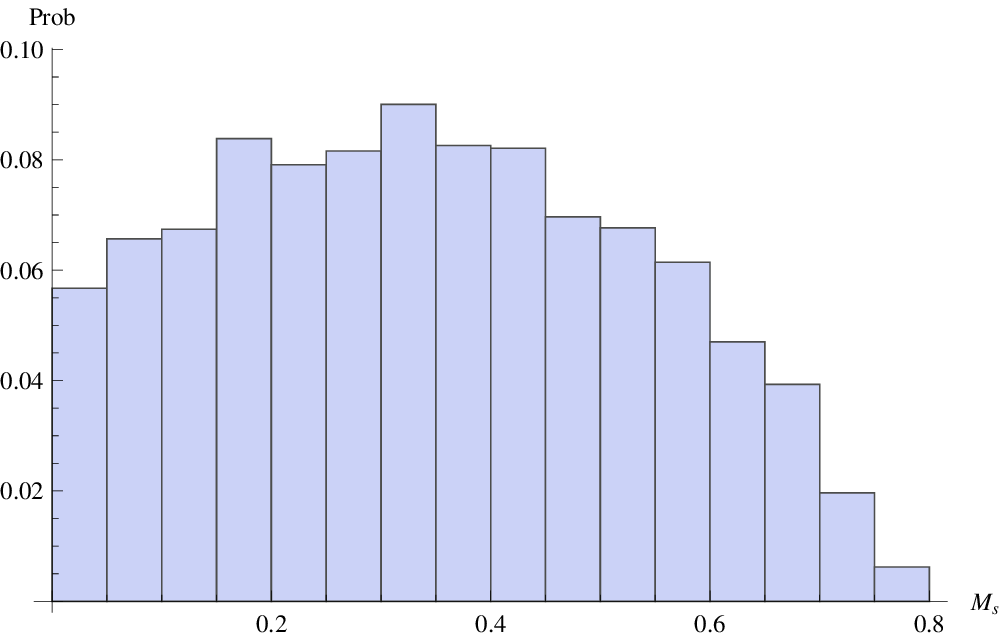}
\caption{The probability distribution  of $\theta$ (Left Panel) and $M_S$(Right Panel) of all viable parameter configurations.
 \label{fig:thetaMs_distribution} }
\end{center}
\end{figure}

The $\bar{\kappa}$ values for  points which successfully stay under the direct search bound turn out to be small comparing
to the electroweak scale $v$, from $ |\bar{\kappa}|\lesssim 0.02$ GeV for $M_\rho< M_H/2$
to $\lesssim 3$ GeV for $M_\rho \sim 2$ TeV, see the left panel of Fig.\ref{fig:kappabar_mchi}.
Our scan shows that it seems to have equal probability to be either positive or negative.
  \begin{figure}
\begin{center}
\includegraphics[width=0.4\textwidth]{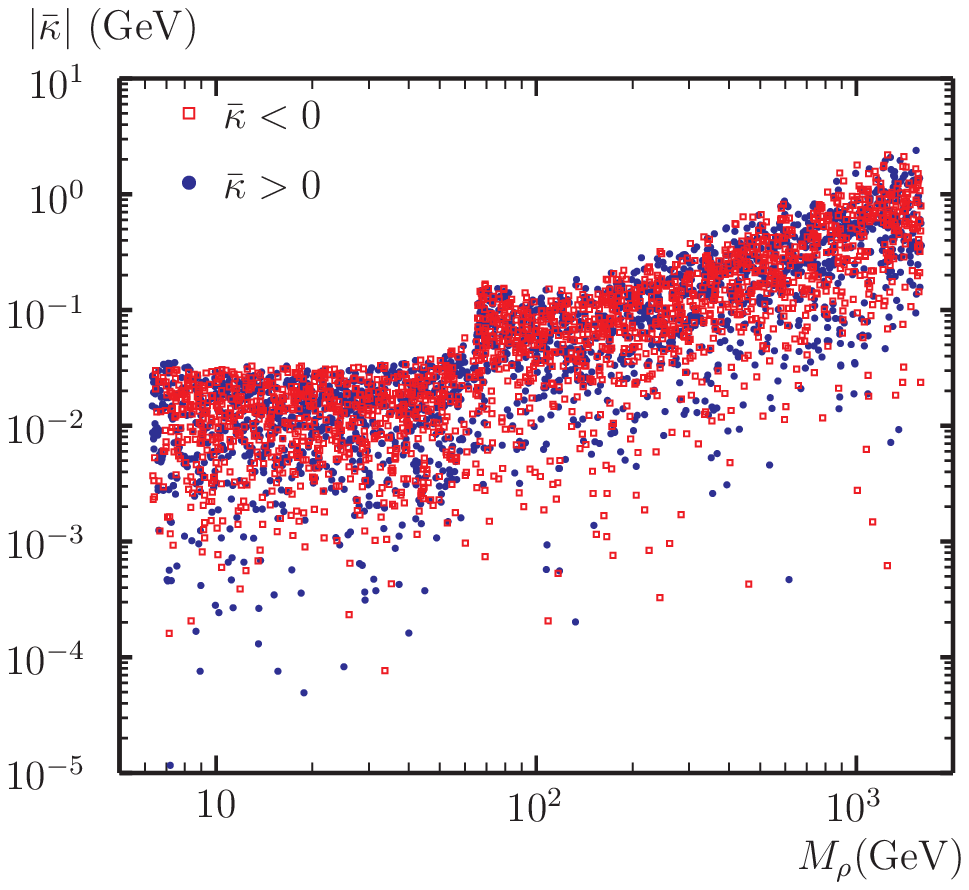}
\includegraphics[width=0.4\textwidth]{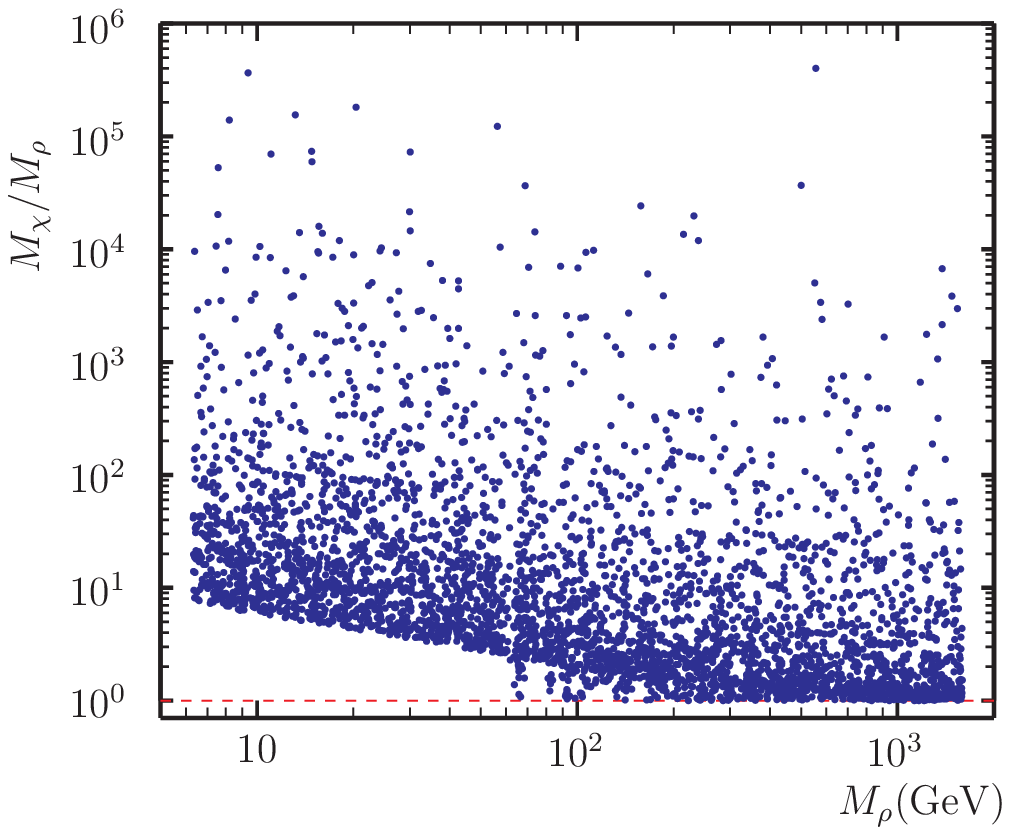}
\caption{Left panel: The distribution of $\bar{\kappa}$ vs $M_\rho$.
Right panel: The mass ratio of two $Z_2$-odd particles vs $M_\rho$
 \label{fig:kappabar_mchi} }
\end{center}
\end{figure}
The mass of the next to lightest $Z_2$-odd particle has a wide range of distribution, see the right panel in Fig.\ref{fig:kappabar_mchi}.
In general,  $M_\chi$  tend to be close to $M_\rho$ at large
$M_\rho (>M_H/2)$, and the mass ratio $M_\chi/M_\rho$ gets larger as $M_\rho$ gets smaller.
The most probable band follows a rough relation $ M_\chi/M_\rho \sim 3 \times (1 \mbox{TeV}/M_\rho)^{1/2}$.
And this result shows that  for DM heavier than $\sim 1$ TeV one also needs to take the coannihilation processes into account.

The distribution of $\lambda_{\Phi H}$ and $\lambda_{\Phi S}$ for different $M_\rho$ are compared in Fig.\ref{fig:LamPH_LamPS}.
The distribution of $\lambda_{\Phi H}$ seems to be symmetric for either sign except at around $M_\rho\sim M_H/2$ where larger value of
negative $\lambda_{\Phi H}$ is preferred over the positive one.
On the other hand, only about $0.03\%$ of successful solutions have negative $\lambda_{\Phi S}$ (red squares in the figure) due to that $\lambda_S \sim (M_S/v_S)^2/2$ is very small which results in a tight lower bound for negative $\lambda_{\Phi S}$.
It is easy to see that the lighter the $\rho$, the smaller  $|\lambda_{\Phi H}|$ and $\lambda_{\Phi S}$. When $M_\rho < M_H/2$, the $\lambda_{\Phi S}$ roughly follows a scaling law that $\lambda_{\Phi S} \propto M_\rho $.
\begin{figure}
\begin{center}
\includegraphics[width=0.42\textwidth]{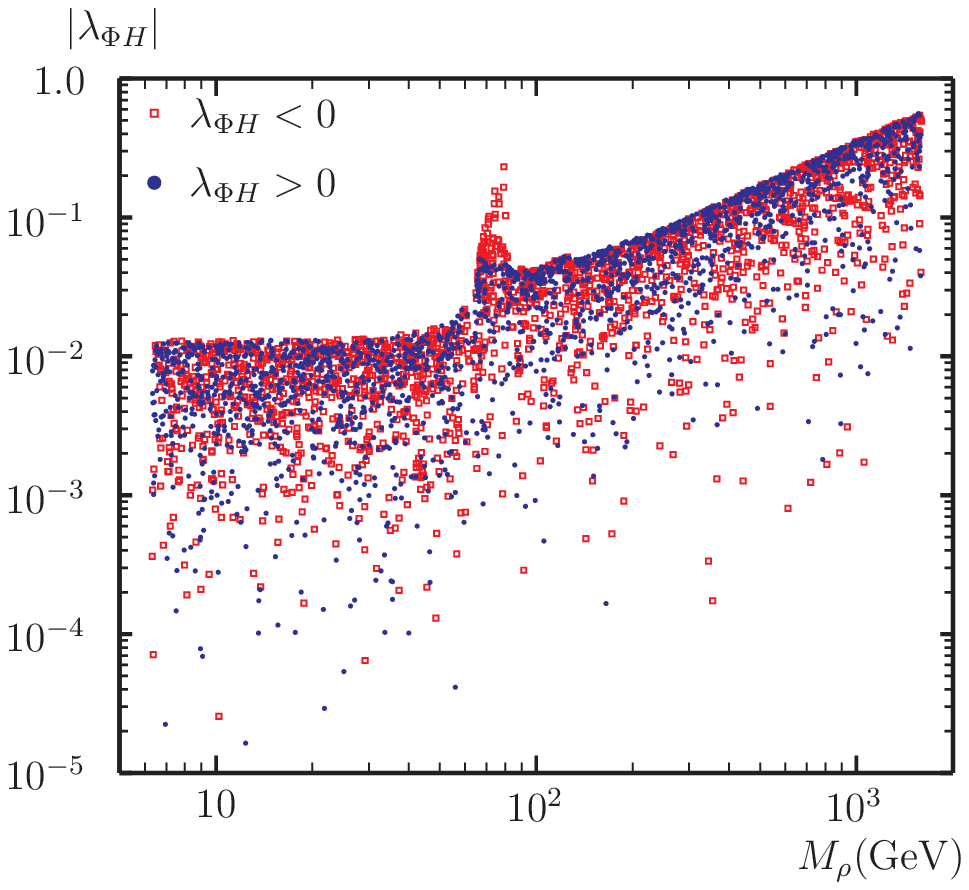}
\includegraphics[width=0.42\textwidth]{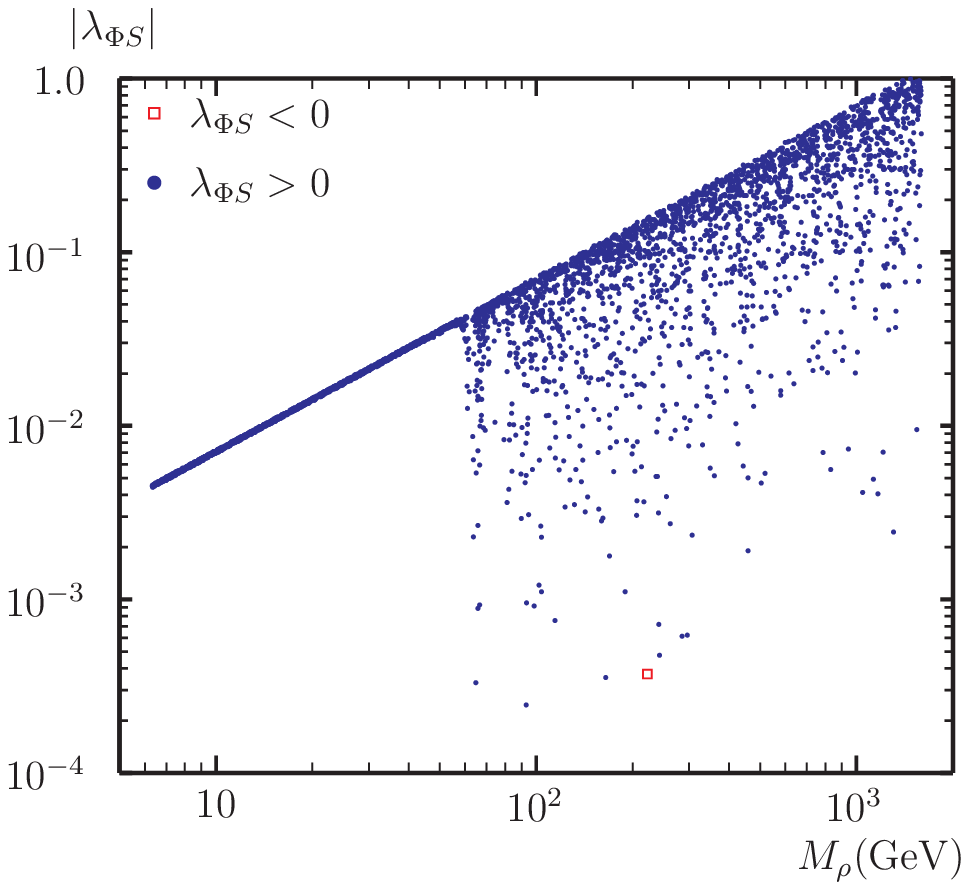}
\caption{ $\lambda_{\phi H}$ (left panel) and $\lambda_{\phi S}$ (right panel) vs  $M_\rho$. \label{fig:LamPH_LamPS} }
\end{center}
\end{figure}

The result of our scan shows that $v_s$ is insensitive to $M_\rho$, see Fig.(\ref{fig:VS}). The lepton number breaking scale generally peaks at around $0.6-3$ TeV and extends to around $10^5$ TeV with monotonically decreasing probability. This puts the right-handed neutrino $N_R$ within reach for LHC searches. However, a detail study  will be needed as the background for heavy neutrinos searches at the LHC is expected to be large or even prohibitive.

\begin{figure}
\begin{center}
\includegraphics[width=0.4\textwidth]{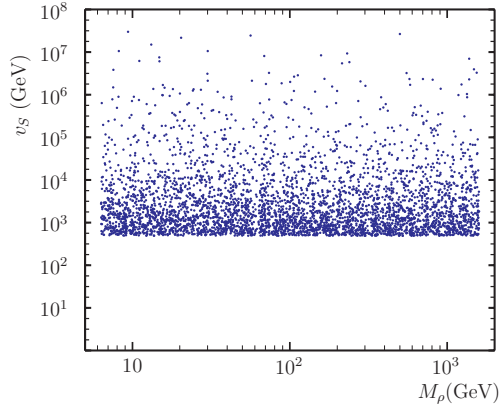}
\caption{The lepton number violating scale $v_S$ vs  $M_\rho$.}
\label{fig:VS}
\end{center}
\end{figure}

In Fig. \ref{fig:sigmaSV}, $\langle \sigma_S v \rangle / \langle \sigma v\rangle_{total}$  and
$\langle \sigma_\omega v \rangle / \langle \sigma v\rangle_{total}$ are  displayed.
It is easy to see that $\rho\rho\ra ss$ is the dominant annihilation channel when $M_\rho<M_H/2$.
On the other hand,  the $\rho\rho\ra \omega\omega$  annihilation channel starts to contribute when  $M_\rho>M_H/2$
with chances to be sizable when $M_\rho$ becomes heavier.
Nevertheless, the $\rho\rho\ra \omega\omega$ channel is usually insignificant in most of the parameter space.
Therefore, when  $M_\rho>M_H/2$,  the processes of dark matter annihilate into $ss$ and the SM particles pair
are the major players determining the thermal dark matter relic density.
\begin{figure}
\begin{center}
\includegraphics[width=0.4\textwidth]{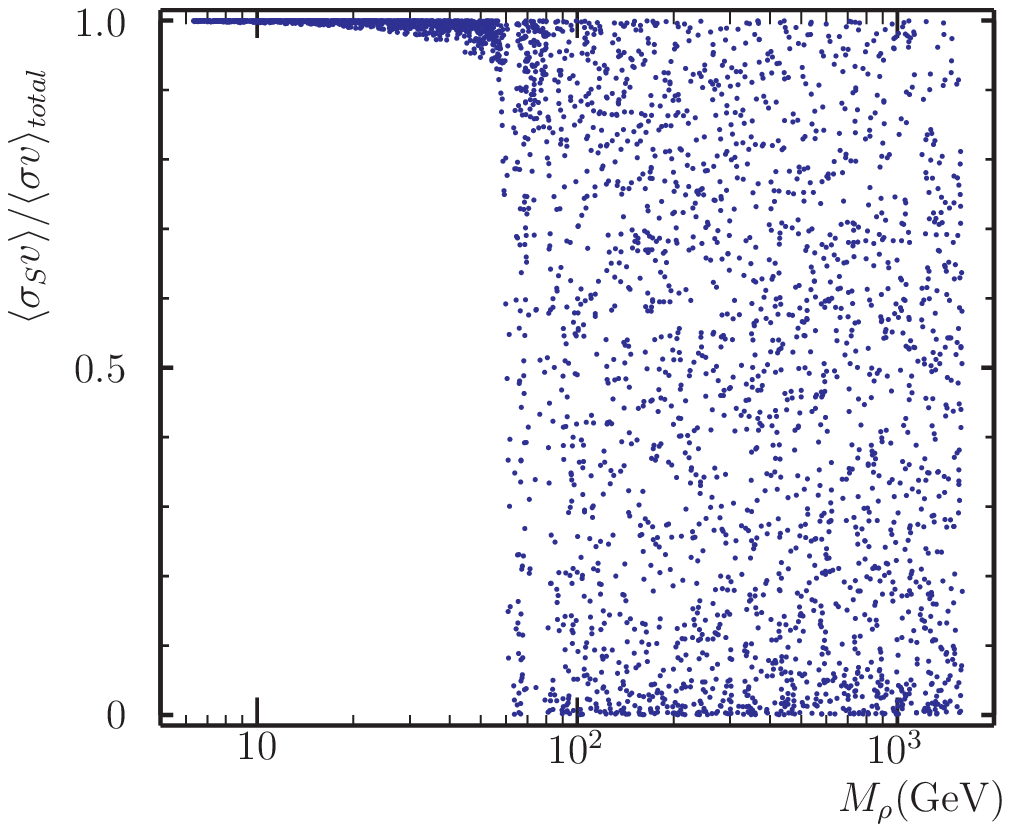}
\includegraphics[width=0.4\textwidth]{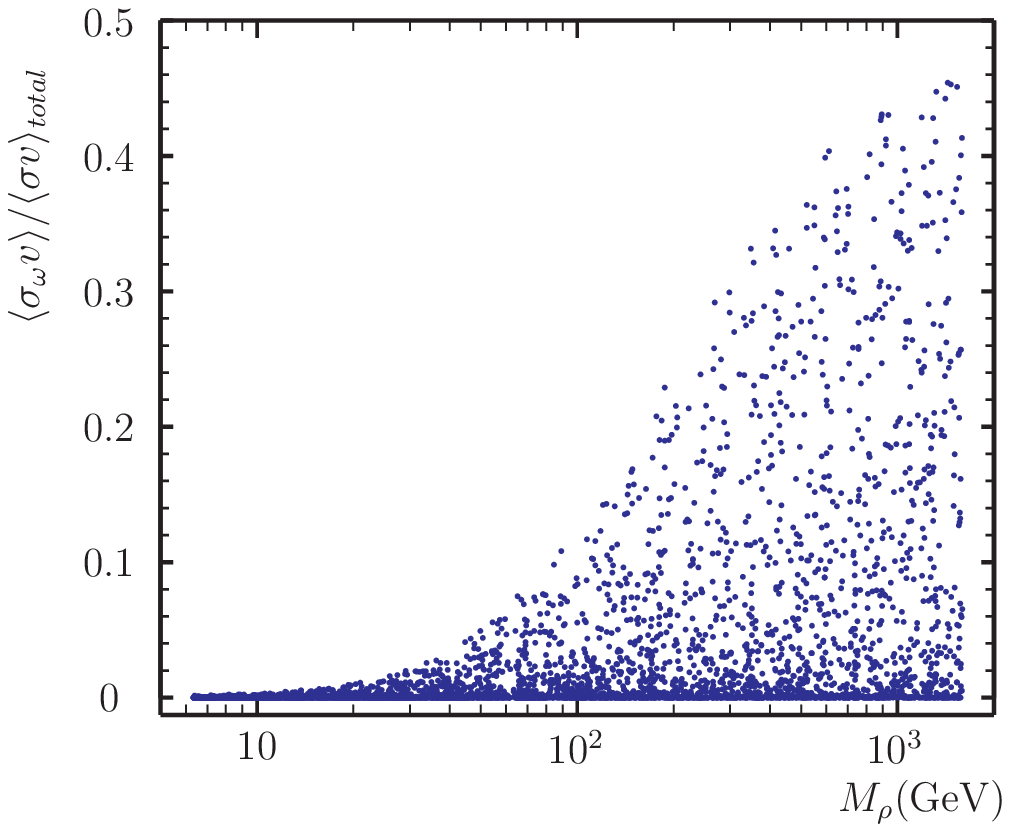}
\caption{ $\langle \sigma_S v \rangle / \langle \sigma v\rangle_{total}$ (left  panel), and  $\langle \sigma_\omega v \rangle / \langle \sigma v\rangle_{total}$(right panel) vs  $M_\rho$ \label{fig:sigmaSV}. }
\end{center}
\end{figure}

Finally, the resulting  spin-independent elastic $\rho$-nucleon scattering cross section v.s. $M_\rho$
is displayed in Fig.\ref{fig:mrho_DD}, where the  LUX $90\%$ confidence limit can be clearly seen.
Most of the data points are within the range between the current LUX limit and one order smaller than the
current limit which can be probed with the LUX 300-day projected sensitivity.
  \begin{figure}
\begin{center}
\includegraphics[width=0.5\textwidth]{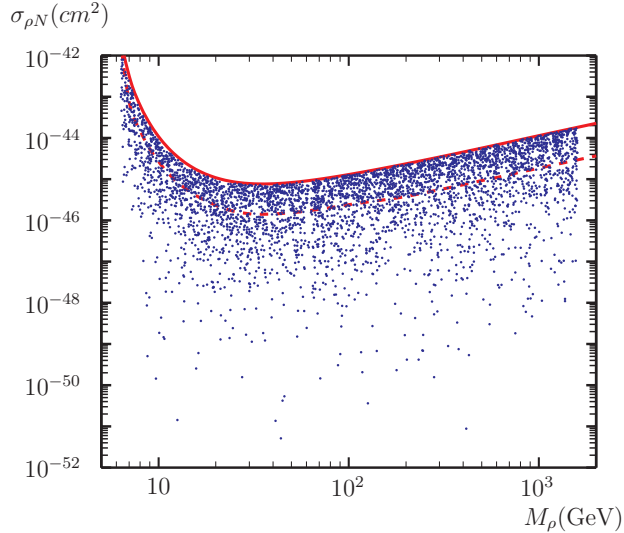}
\caption{Spin-independent elastic $\rho$-nucleon scattering cross section v.s. $M_\rho$.
Where the solid line is the current LUX limit and the dashed line is the LUX 300-day projected sensitivity.
 \label{fig:mrho_DD} }
\end{center}
\end{figure}

\section{Gamma Rays from the Galactic Center}
A recent study indicates that the low-energy ($\sim 1-3$ GeV) gamma-ray excess at the Galactic center
can be accommodated by a $30-40$ GeV dark matter particle annihilating into $b\bar{b}$ with
an annihilation cross section of $\langle \sigma v \rangle =(1.4-2.0)\times 10^{-26} cm^3/s$\cite{Daylan:2014rsa}.
In this section we shall see whether the low-energy gamma-ray excess at the Galactic center can be
accommodated with $M_\rho$ $\sim 30-40$ GeV in our model { (see \cite{Martin:2014sxa} for a discussion on other scenarios with the same kinematics.)
As discussed in previous section,  $\rho\rho\ra ss$ will then be the dominant annihilation channel.
Since the model predicts that the light scalar $s$ has mass $M_s<1$ GeV, it can only decay into light quarks or gluons at the parton level.
Thus, we will discuss and compare the gamma ray spectrum generated from the decay of $s$ with energy $\sim  30-40 \ \mbox{GeV}$
to that of the benchmark scenario in \cite{Daylan:2014rsa}.

The gamma-ray spectrum produced from dark matter annihilating into $f\bar{f}$  is given by
\beq\label{eq:GCgammaspectrum}
{ d\Phi_\gamma \over d \Omega dE_\gamma}
=\sum_i \frac{d N_\gamma^i}{d E_\gamma} { \left\langle \sigma_i v\right\rangle\over 4\pi M_{DM}^2}
\times \left[ B \times \int_{\mbox{line of sight}} \rho_{DM}^2 dl\right]\,,
\eeq
where $i$ represents the final state particle specie, $d\Omega$ is the solid angle seen from the earth,  $\rho_{DM}$ is the  dark matter mass density, and the boost factor is defined as $B\equiv \langle \rho_{DM}^2\rangle/ \langle\rho_{DM}\rangle^2 $.
The  booster factor is close to its minimum $=1.0$ when  fluctuation of the Galactic dark matter mass density  is small.
If there is only one kind of dark matter,  the dark matter number density will be  $\rho_{DM} / M_{DM}$ and that explains the $M_{DM}^2$ factor in the denominator.
In the square bracket,  the boost factor and the $\rho_{DM}^2$ integral along the line of sight are purely astronomical and strongly model dependent.
Here $\frac{d N_\gamma^i}{d E_\gamma}$ is the gamma ray spectrum produced by the energetic  quarks or $W/Z$ boson with initial energy $E_i=M_{DM}$ which hadronizes into $\pi^0$ and other mesons and they  decay
into photons subsequently.
With the same initial energy, the top and bottom pairs yield the softer gamma rays, and light quark or gluon pairs yield  the harder gamma rays. The gamma-ray spectrum produced by $W$ and $Z$ is in between the spectrum from the light quark and heavy quark.
This function can only be fitted from experiments and have been encoded into many computer programs.
For a ballpark estimation, we adopt a simple approximation proposed by\cite{gammaray_spectrum}:
\beq
\label{eq:gamma_function}
\frac{d N_\gamma^i}{d E_\gamma} \sim {a_i (M_{DM})^{0.5} \over (E_\gamma)^{1.5}}
\times e^{- b_i E_\gamma /M_{DM}}\,,
\eeq
with $(a,b)= \{(1.0,10.7),(1.1,15.1),(0.95,6.5),(0.73,7.76)\}$
for $i=\{ b\bar{b}, t\bar{t}, u\bar{u}, W^+W^-/Z Z \}$.
We shall make use of this approximation and estimate the gamma rays produced from  $\rho \rho \ra ss$, where $s$ subsequently decays into light quarks or gluons (so we take $(a,b)= (0.95,6.5)$ ).

From the rest frame of $s$, we boost the isotropically distributed $s\ra q\bar{q}, gg$ to energy $E_s=M_\rho$
so that the energy carried by quark or gluon in the dark matter annihilation center-of-mass (c.m.) frame is in the range between
 $E_f^{min}=(M_\rho/ 2)[ 1- \sqrt{1 -(M_s/M_\rho)^2}\sqrt{1- (2 m_f/M_s)^2} ]$
 and  $E_f^{max}=(M_\rho/ 2)[ 1+ \sqrt{1 -(M_s/M_\rho)^2}\sqrt{1- (2 m_f/M_s)^2} ]$, where $m_f$ is the mass of quark and $0$ for gluon.
After averaging over all possible direction,  we obtain the following normalized differential probability of finding a light quark or gluon with energy $E_f$ in the c.m. frame:
\beq
{ d P_s \over d E_f} =\frac{4}{\pi M_\rho}
{ \sqrt{\left(1-\frac{M_s^2}{M_\rho^2}\right)\left(1- \frac{4 m_f^2}{M_s^2}\right) -\left(1-\frac{2 E_f}{M_\rho}\right)^2} \over
\left(1-\frac{m_S^2}{M_\rho^2}\right)\left(1- \frac{4 m_f^2}{M_s^2}\right) }\,,
\eeq
which peaks at $E_f=M_\rho/2$ and smoothly drops to zero at $E_f^{max}$ and $E_f^{min}$.
Notice that $E_f$ can be viewed as the dark matter with an effective mass
$M^{eff}_{DM}=E_f$  annihilating into $f\bar{f}$. Hence, one can convolute this distribution with the photon spectrum function and the contribution to the gamma ray spectrum
from $\rho\rho\ra ss$ can be expressed as
\beq
{d \Phi_\gamma \over d \Omega d E_\gamma}
=\left[ \int^{E_f^{max}}_{E_f^{min}} d E_f \frac{a_s (E_f)^{0.5} }{(E_\gamma)^{1.5}}
e^{- \frac{b_s E_\gamma}{E_f} } \left(2\frac{d P_s}{d E_f}\right) \right]
\times { Br_h \left\langle \sigma_S v \right\rangle \times B \over 4\pi M_\rho^2}
\int_{l.o.s.} \rho_{DM}^2 dl\,,
\eeq
where $Br_h$ is the hadronic decay branching ratio of $s$. The factor $2$ associated the differential probability
 is to account for that there are 4 final light quarks or gluons from the two decaying $s$.
We will use Eq.(\ref{eq:s_decay_ratio}) to approximate $Br_h \left\langle \sigma_S v \right\rangle $.
Since Eq.(\ref{eq:GCgammaspectrum}) can be factorized into an astrophysical part and the particle physics part,
we concentrate on the particle physics component only and adopt the best fit from \cite{Daylan:2014rsa}.
We use Eq.(\ref{eq:gamma_function}) for the gamma-ray spectrum from a dark matter of mass $35$ GeV with an annihilating cross section into $b\bar{b}$ of $1.42\times 10^{-9} (\text{GeV})^{-2}$ as the benchmark.
We found that in our model $M_\rho = 37.30$ GeV and $B\times Br_h=0.507$ give the best fit to the benchmark spectrum
between energy $0.3-30$ GeV where we equally divide the energy logarithm  into 12 bins and the relative
uncertainty is about $3\%$ for each data point, see Fig.7 in \cite{Daylan:2014rsa}.
On the other hand, the best fit of  our model has  $M_\rho = 36.53$ GeV and $B\times Br_h=0.499$ if we try to best match
the benchmark spectrum between a narrower range $0.3-10.0$ GeV, see Fig.\ref{fig:bestFIT} for the comparisons.
   \begin{figure}
\begin{center}
\includegraphics[width=0.6\textwidth]{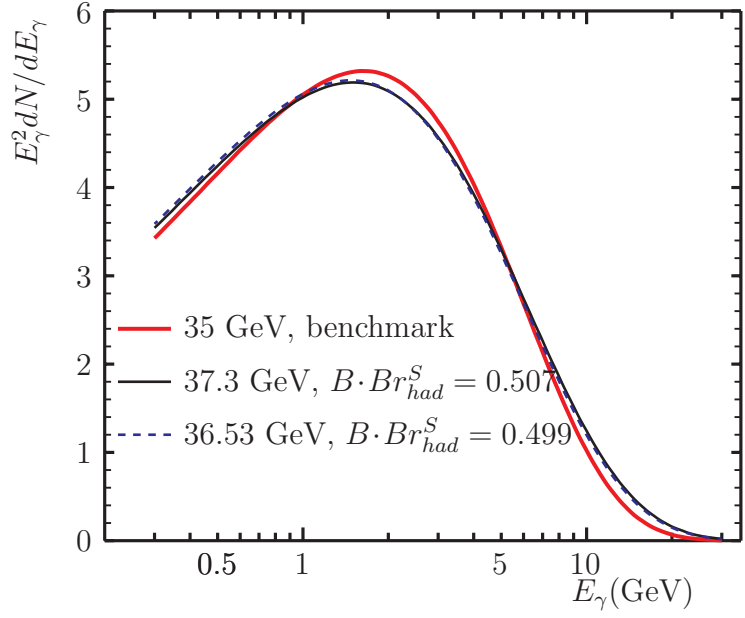}
\caption{Comparison of the  diffuse gamma-ray spectrum $E^2 dN/dE $ with arbitrary unit from
the benchmark $35$GeV DM and $\langle \sigma_b v\rangle =1.42 \times 10^{-9}(GeV)^{-2}$ (red),
$M_\rho=37.30$ GeV and $B\times Br_h=0.507$ in our model(black), and
$M_\rho=36.53$ GeV and $B\times Br_h=0.499$ in our model(blue dash).
 \label{fig:bestFIT} }
\end{center}
\end{figure}
Our best fit has a slightly harder spectrum at $E_\gamma =10$ GeV, which is actually better than the
benchmark spectrum, see Fig.7 in \cite{Daylan:2014rsa}.

With the target range set, we zoom in our numerical search and focus on the points with
$M_\rho= 37.0\pm 5.0$ GeV.  For  $M_s > (M_K-m_\pi)$, the  $B\ra K s \ra K \mu\bar{\mu}$ and $B\ra K s \ra K + (\mbox{nothing})$  experiments
constrain $\theta$ to be $\lesssim 0.01 $ \cite{BKs}, and the corresponding  boost factor  ranges from $\sim 10$ to $2000$, depending on the mixing, see Fig.\ref{fig:BoostF}.

\begin{figure}
\begin{center}
\includegraphics[width=0.5\textwidth]{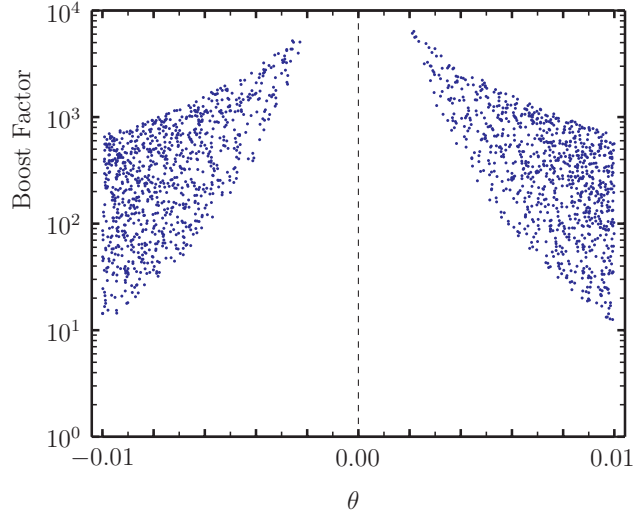}
\caption{ The boost factor needed for $m_S> M_K-m_\pi$  to get the best-fit benchmark as a function of $\theta$.}
\label{fig:BoostF}
\end{center}
\end{figure}
If a smaller boost factor is preferred, it looks like the model needs to be stretched to simultaneously meet the rare B decay limits and the Galactic center gamma ray excess. However, the boost factor is very sensitive to  the decoupling temperature, $T_{dec}$, see Eq.(\ref{eq:gb2DR}).
For example, if the decoupling temperature is slightly raised to $T_{dec}\sim 2.2 m_\mu$ from $2.0 m_\mu$, the smallest $B$ at
$|\theta| \sim 0.01$ can be pushed down to $\sim 6.0$ from $\sim 13.0$ for $T_{dec}\sim 2.0 m_\mu$.
Given the large uncertainties in astrophysics, cosmology, and hadronic form factors, our model can accommodate the Galactic center gamma-ray with a $\sim 40$ GeV $\rho$ and satisfy the B-decay limits at the same time.
However, the constraints are quite tight and a very small mixing and a higher decoupling temperature are preferred if the model is to fit the gamma rays from the Galactic
center data as it stands now.

\section {Rare Higgs Boson Decays}
Our model belongs to the category of Higgs portal models \cite{HP,shadow} in which the dark sector communicates with the SM via Higgs couplings only. The characteristic
signatures are new rare Higgs decays. The new decay modes comes from (a) $h\ra \omega \omega $ and (b) $ h \ra \rho\rho$ and (c) $h \ra  s s $. All three channels will give rise to $\Gamma_{\text{inv}}$ for the Higgs with $s$ decaying into a pair of $\omega$'s. In our parameter scan we require the invisible decay to concur with the experimental limit. The scatter plot given in Fig.\ref{fig: Hinv} shows the preferred values.
$\Gamma_{\text{inv}}$ is clearly divided into two regions with the boundary at around $M_\rho=M_H/2$.
 For $M_\rho> M_H/2$, the model prefers a small invisible decay width, on the other hand,  $\Gamma_{\text{inv}}$
 could be as large as the input limit, $0.8$ MeV, for $M_\rho<M_H/2 $.
\begin{figure}
\begin{center}
\includegraphics[width=0.5\textwidth]{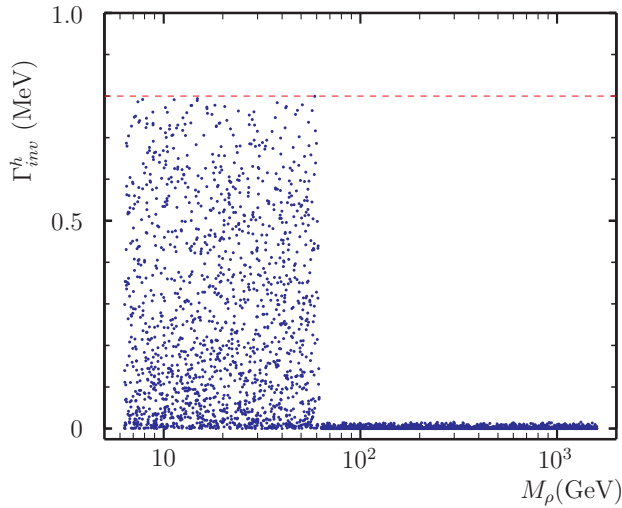}
\caption{ SM Higgs invisible decay width $\Gamma^H_{inv}$ as a function of$m_\rho$.}
\label{fig: Hinv}
\end{center}
\end{figure}
Although this not a robust prediction it can be used as a check if other signals are seen.

Other interesting Higgs boson decays comes from process-(c) and $s$ subsequently decays into lepton pairs, hadrons or two photons depending on its mass. Since $s$ is light and have a long lifetime displaced vertices is a clear possibility. Previously this interesting signal is considered in the context of supersymmetric models \cite{SZ},  leptoquark
models \cite{KNSW}, and heavy neutrino searches \cite{HK}.
 Here the displaced vertices originates from  the Higgs boson decays. In Fig.\ref{fig:S_decay}, data points are displayed to show the  decay width and invisible decay branching ratio for $s\ra \omega \omega$ as a function of $M_s$.
\begin{figure}
\begin{center}
\includegraphics[width=0.45\textwidth]{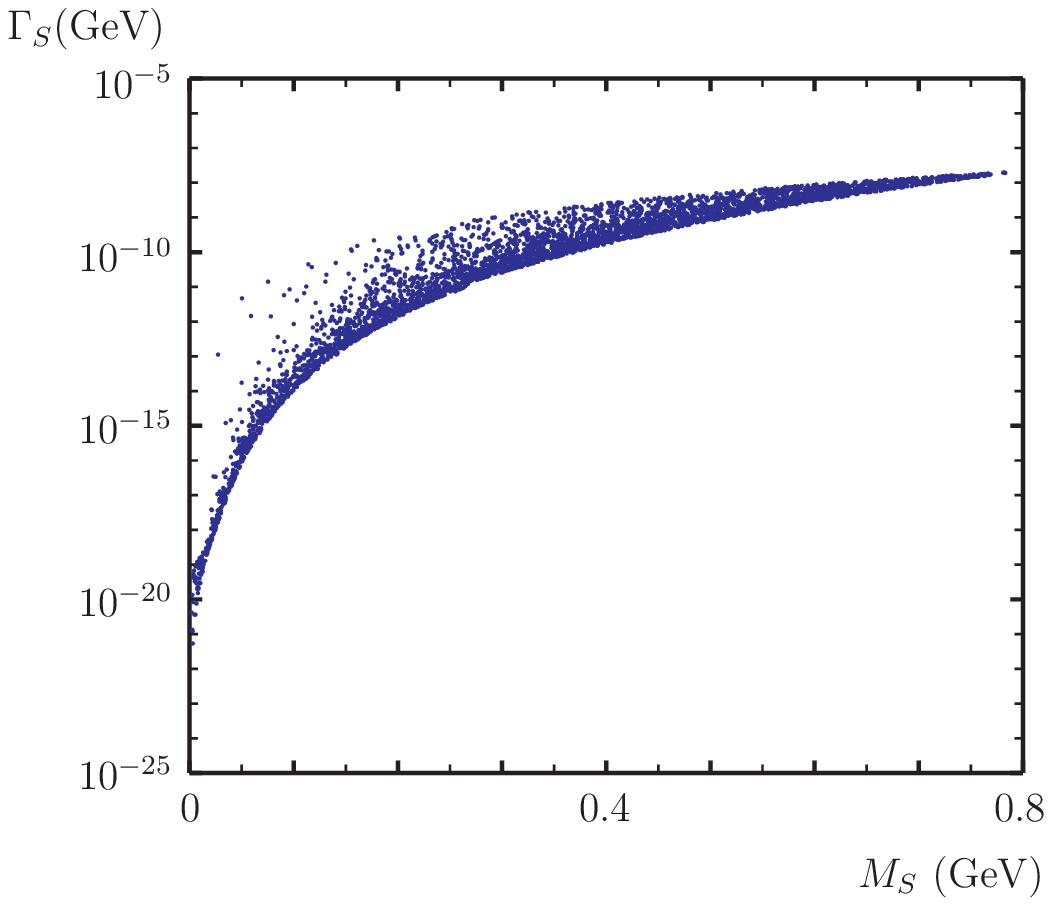}
\includegraphics[width=0.45\textwidth]{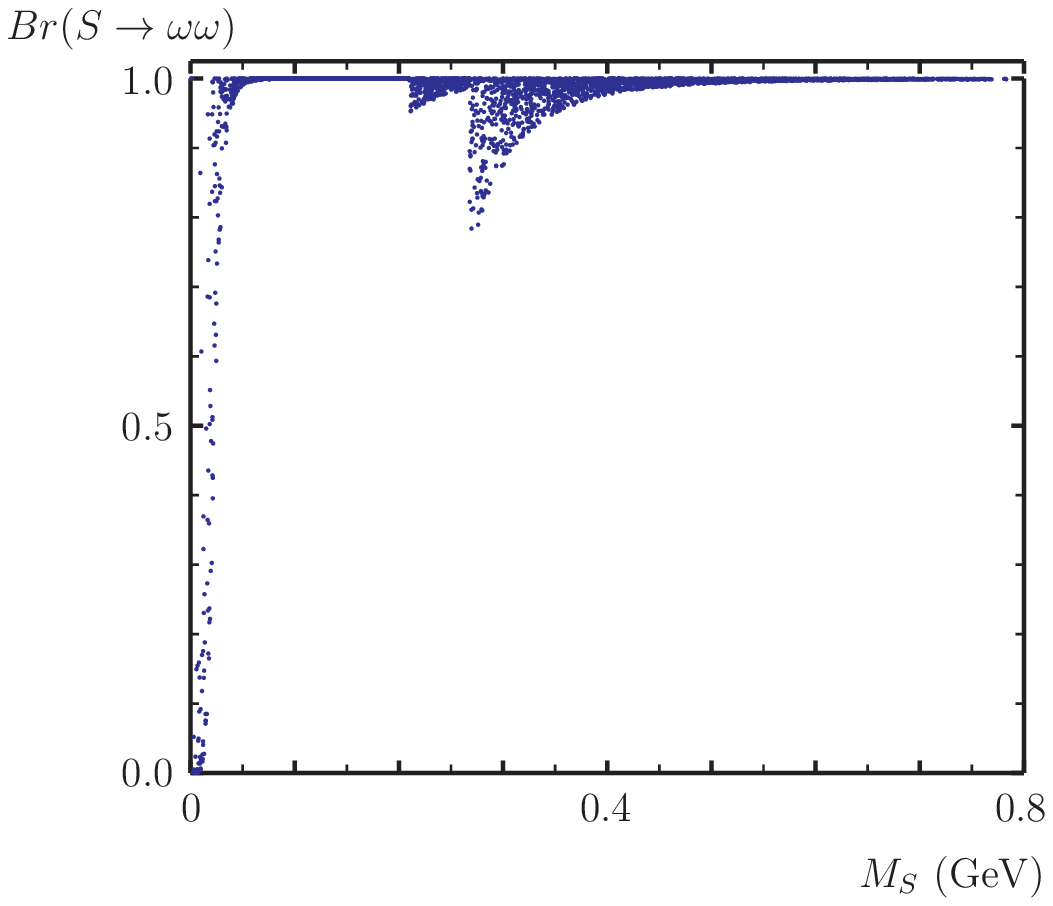}
\caption{Left panel: light scalar $s$ decay width  vs $M_s$, Right panel:  invisible decay branching ratio vs   $M_s$.
  \label{fig:S_decay} }
\end{center}
\end{figure}
Both the decay width and invisible decay branching ratio of $s$ are quite independent of the dark matter mass, $M_\rho$.
 This is because the $s$ decays can be completely determined by $\theta$, $M_s$, and $v_s$.
 One can see the jumps in $Br(s\ra \omega\omega)$ at the
$M_s = 2 m_\mu$ and $M_s=2 M_\pi$ thresholds.
The corresponding decay vertex displacement ranges from $\sim (M_H/2 M_s)\times 10^{-5}$ cm for $M_s \sim 0.8$ GeV to $\sim (M_H/2 M_s)\times  1$ cm for $M_s\sim 0.1$ GeV (see Fig.\ref{fig:ctau}).
\begin{figure}
\begin{center}
\includegraphics[width=0.6\textwidth]{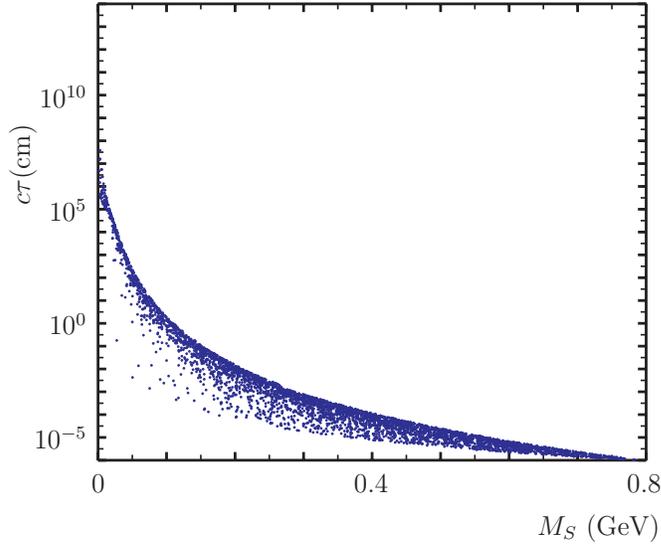}
\caption{Expected $c\tau$ (in cm) as a function of $m_s$.}
\label{fig:ctau}
\end{center}
\end{figure}
The $s\ra e^+ e^-$ and $s\ra 2\gamma$ decays
are almost always overwhelmed by the $s\ra \omega \omega$ channel.

For $2M_\pi < M_s < 1 {\text {GeV}}$ the Higgs boson can have spectacular decays from the chain $ h\ra s \ s \ra \pi\pi (\mu\mu)+ \pi\pi (\mu\mu)$ or $h\ra s \ s \ra \pi\pi (\mu\mu) + \slashed{E}$ and the missing energy $\slashed{E}$ originates from one $s$ decaying into $\omega\omega$ and hence is recoiling against the pair of detected particles. We did not include kaon modes since they are kinematically suppressed.

The  $h\ra 4\mu$ is particularly interesting and has been searched  for by the CMS Collaboration \cite{CMS4mu}.
The cross section is given by
\beq
\sigma(h\ra 4 \mu)=\sigma(h)Br(h\ra s s)\left[Br(s\ra \mu \mu)\right]^2\,.
\eeq
For $2m_\mu < M_s < 2 m_\pi$,  $Br(s\ra \mu\mu)\lesssim 0.05$ and the CMS limit of $\sigma(h\ra4\mu) < 0.86$ fb at 95\% C.L. implies $\Gamma(h\ra s s) \lesssim 9.3 \times 10^{-2} \ {\text{MeV}}$ if the values $\sigma(h)=15.13$ pb and $\Gamma_H = 4.07$ MeV are used.
For $2m_\pi < M_s < 1$ GeV, ${\text{Br}}(s\ra \mu\mu) \approx 0.01$ and we have instead $\Gamma(h\ra s s) \lesssim 2.3 $ MeV.
Basically  the current CMS limit post no constraint on this model. This is shown in Fig.\ref{fig:SMwidth}.
\begin{figure}
\begin{center}
\includegraphics[width=.45\textwidth]{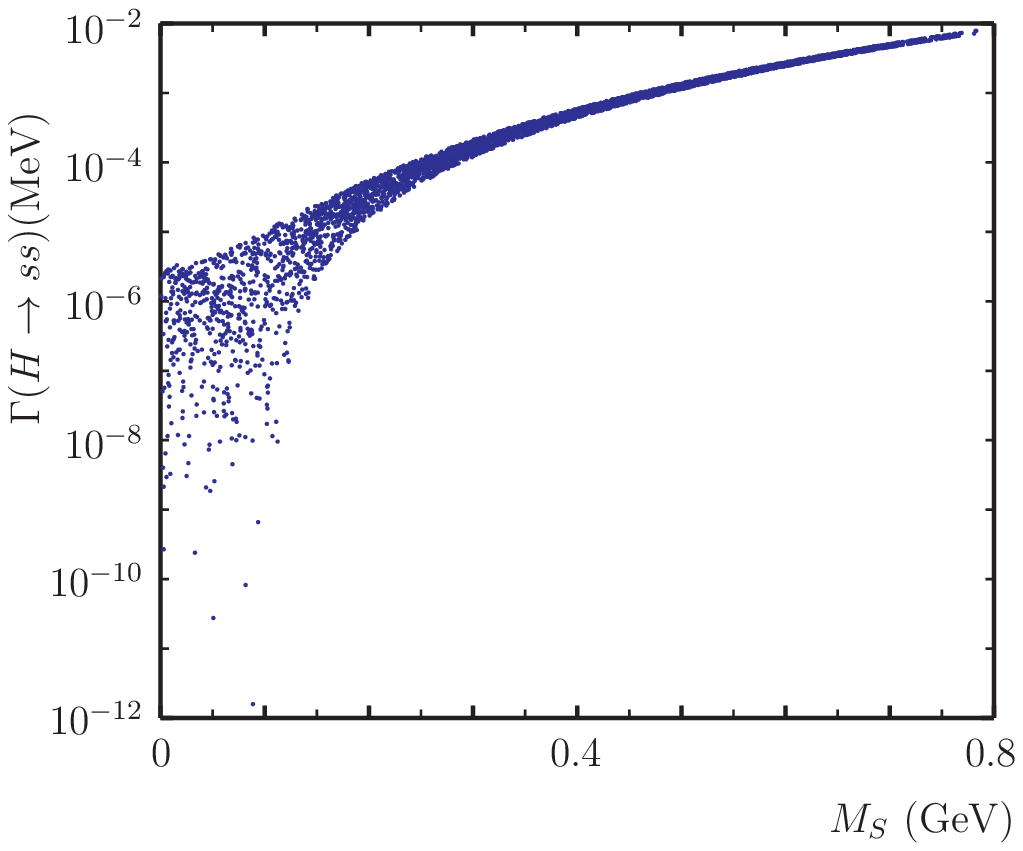}
\includegraphics[width=.45\textwidth]{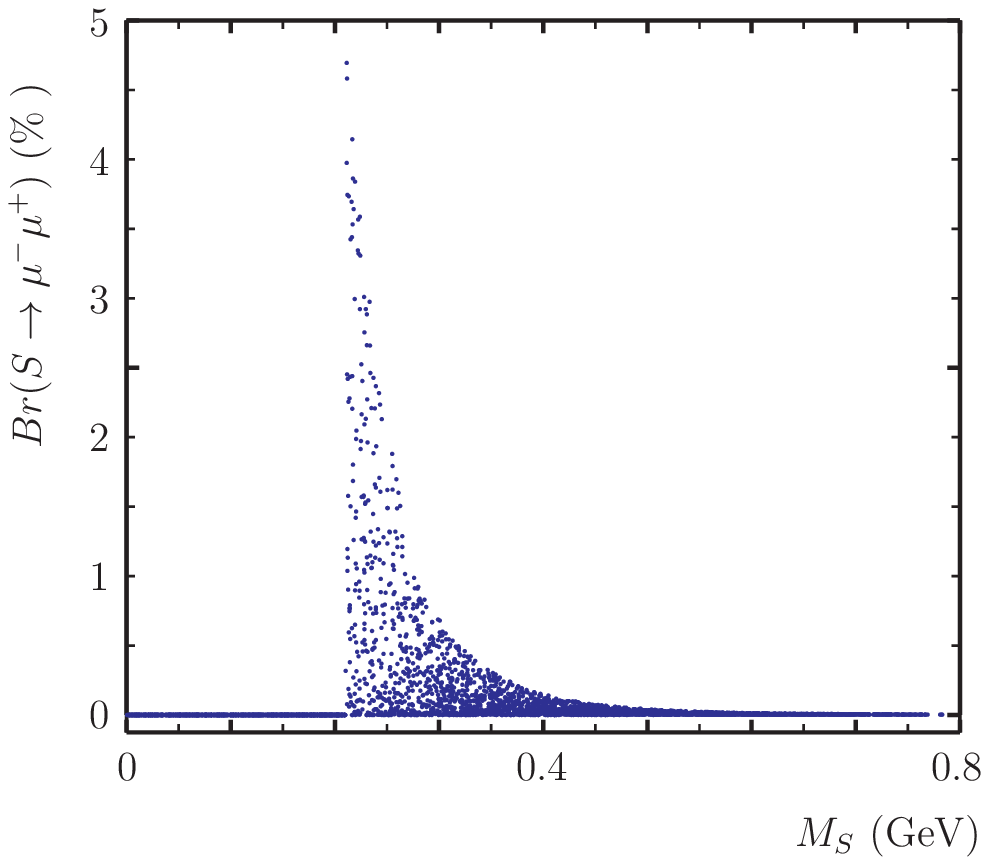}
\caption{Decay width $\Gamma(h\ra ss)$ in MeV (left panel) and ${\text{Br}}(s\ra \mu\mu)$(right panel).}
\label{fig:SMwidth}
\end{center}
\end{figure}
Interestingly the cross section for $2 \mu +\slashed{E}$ is almost two orders of magnitude larger  than that for $4 \mu$ by virtue of the larger invisible $s$ branching
ratio and this is given by
\beq
\sigma(h\ra 2\mu +\slashed{E})=2 \sigma(h)Br(h\ra s s)Br(s\ra \mu\mu)Br(s\ra \omega\omega).
\eeq
The prediction of our model is displayed in Fig.\ref{fig:H_mm_ME}.
\begin{figure}
\begin{center}
\includegraphics[width=.45\textwidth]{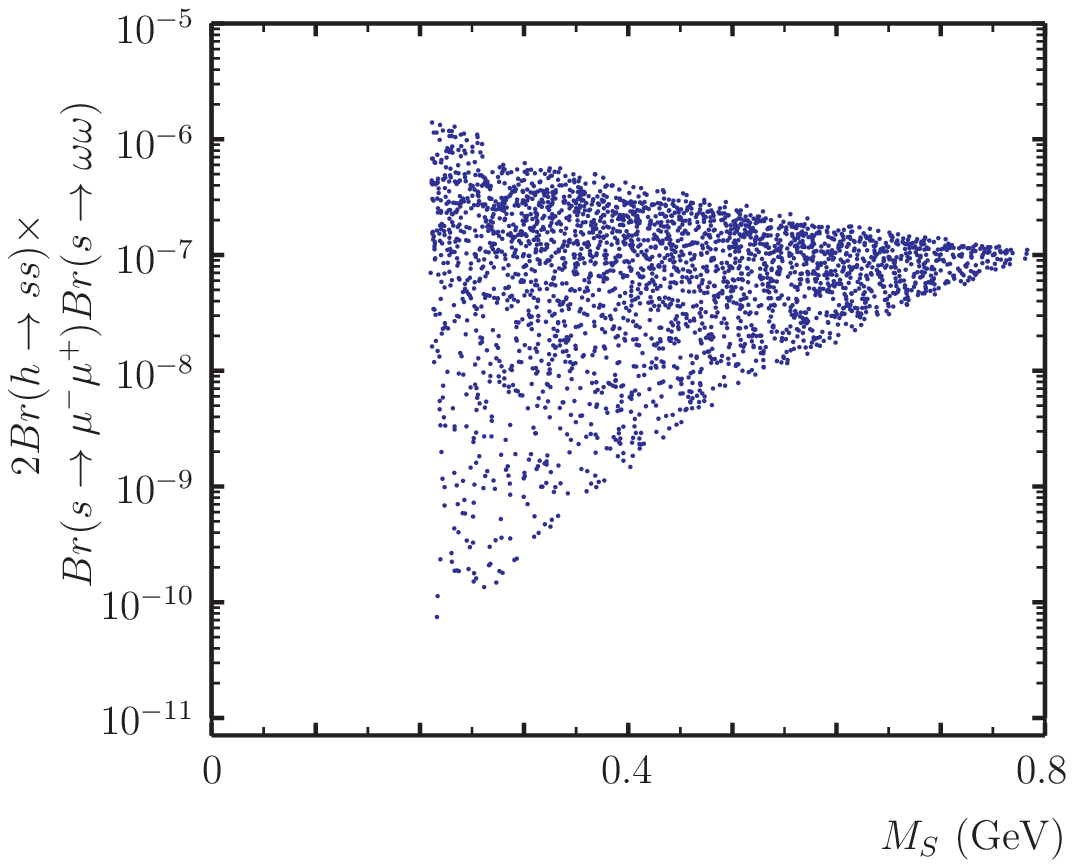}
\includegraphics[width=.45\textwidth]{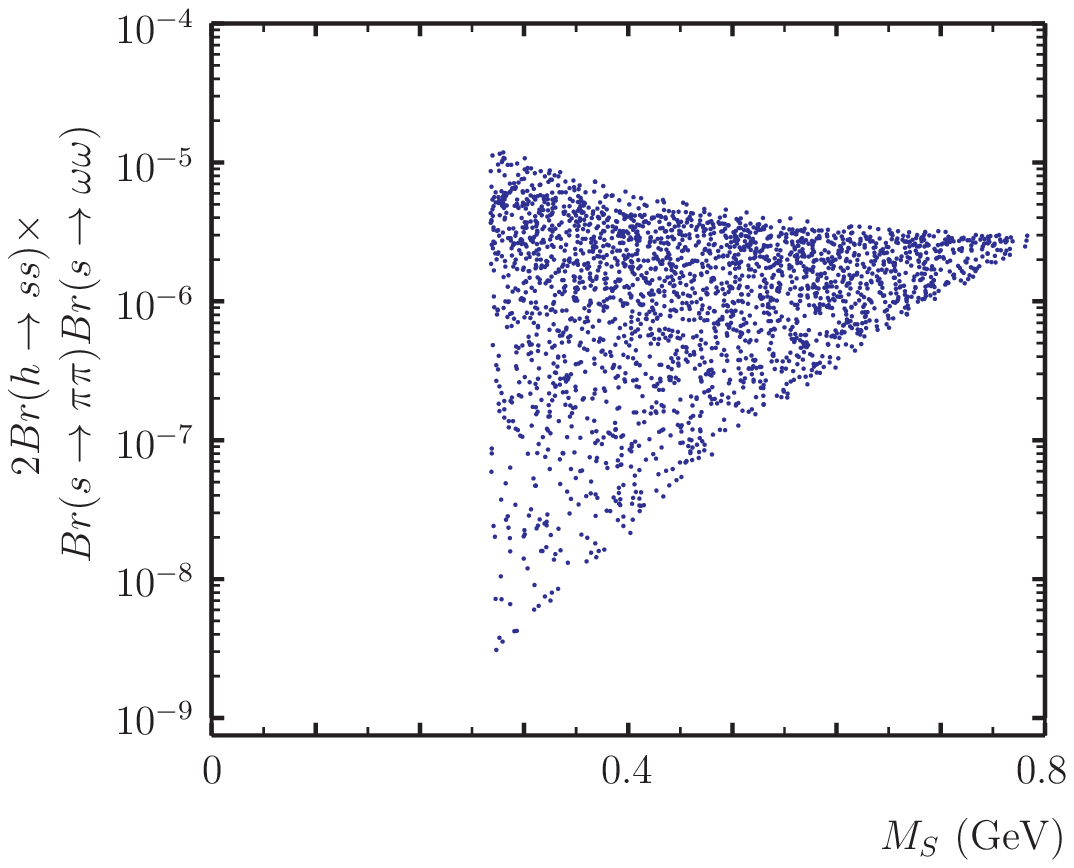}
\caption{ The branching ratios  for  $2 \mu +\slashed{E}$ (Left panel) and  $2 \pi +\slashed{E}$ (right panel)
in our model.}
\label{fig:H_mm_ME}
\end{center}
\end{figure}
The largest branching ratios are around $10^{-6}$ and $10^{-5}$  for  $2 \mu +\slashed{E}$
and  $2 \pi +\slashed{E}$, respectively. For LHC14 the gluon fusion Higgs production cross section is $\sim 50$ pb and hence the high luminosity option will give the necessary rates.
However, the background is expected to be large.
Better signal selection triggers will greatly improve the odds. Detail studies are beyond the scope of this paper.
All these modes can also be  searched for at the ILC or an $e^+e^-$ Higgs factory where the background is much smaller and the events are cleaner.
However, the Higgs production cross section at the $e^+e^-$ machine is roughly 2 orders of magnitude smaller than that at the LHC.
To see these rare Higgs decays at an $e^+e^-$ collider, one needs  $\sim 100$ times larger luminosity than the currently envisioned for these machines.\footnote{ We thank Jessie Shelton for pointing out to us our earlier overly optimistic  estimation. }

\section {Conclusions}
We have augmented the minimal Majoron model for dark radiation with a SM singlet scalar endowed with unit lepton number. The spontaneous breaking of
the global $U(1)_L$ has three important consequences: (1) the Goldstone Majoron can serve as DR, (2) the type-I seesaw mechanism for light neutrino masses
can be implemented, and (3) a $Z_2$ dark parity naturally occurs as a residual symmetry.  The  existence of a stable scalar dark matter is thus natural. Since the new physics
introduced is in the form of SM singlet scalars, they interact with the SM fields with the Higgs boson as the mediator. In order to obtain an acceptable value for
$\Delta N_{\text{eff}}$, that characterizes DR, it is found that the Majoron must decouple at temperature around $m_\mu$ although the exact value is not predicted.
This leads to the existence of a light scalar $s$ that mixes with the SM Higgs boson. In turn it results in spectacular rare Higgs boson decays such as displaced vertices and
muon pairs with missing energy recoiling against the pair themselves. These can be searched for at the LHC. The invisible width of the Higgs boson is also enhanced which perhaps
is best measured at an $e^+e^-$ Higgs factory.
Our numerical analysis also reveals that the lepton number violating scale $v_s$ is in the range of $1 \mbox{TeV } <v_s < 100  \mbox{TeV}$. This gives
additional motivation to search for heavy neutrinos at the LHC.  Again, a TeV $e^+ e^-$ colliders such as CLIC will be more suitable. Certainly we are
 encouraged that the seesaw scale is not hopelessly out of reach.

We have also investigated whether the model can accommodate the reported gamma ray excess from the Galactic center. This can come from $\rho \rho \ra s s$ followed by
$s$ decaying into light hadrons. We found that a $\sim 40 \mbox{GeV} \rho$ can be made consistent with the data. However, tension with rare B meson decays is also present. This can be resolved by making the mixing of $s$ and the Higgs boson very small and also increasing the decoupling temperature.

In conclusion we constructed a minimal model of Majoron dark radiation with a scalar dark matter that satisfies all experimental constraints.
It also has interesting Higgs phenomenology that can be pursued at the high luminosity LHC and  a future super $e^+ e^-$ Higgs factory.

\begin{acknowledgments}
 W.F.C. was supported by the Taiwan MOST under
Grant No.\ 102-2112-M-007-014-MY3. J.N.N is partially supported by the NSERC of Canada.
\end{acknowledgments}


\begin{thebibliography}{99}
\bibitem{Pneff}
G.~Mangano, G.~Miele, S.~Pastor, T.~Pinto, O.~Pisanti and P.D.~ Serpico, Nucl. Phys. {\bf {B729}} 221 (2005).

\bibitem{Planck}
P.~A.~R.~Abe {\it {et al}} (Planck Collaboration) [arXiv. 1303.5076 [astro-ph.CO]].

\bibitem{HST}
A.~G.~Riess {\it{et al}}, Astrophys. J. {\bf 730}, 119 (2011). [Erratum-ibid {\bf 732}, 129(2011)
[arXiv. 1103.2976 [astro-ph.CO]].

\bibitem{W9}
G.~Hinshaw {\it{et al}} [ arXiv. 1212.5226 [astro-ph]].

\bibitem{ACT}
J.~L.~Sievers {\it{et al}} arXiv 1301.0824 [astro-ph.CO]].

\bibitem{SPT}
Z.~Hou {\it{et al}} [ arXiv. 1212.6267 [astro-ph.CO].

\bibitem{Weinberg}
S.~Weinberg,
Phys. Rev. Lett.  {\bf 110}, 241301 (2013) [arXiv:1305.1971 [astro-ph.CO]].

\bibitem{CNW14}
 W.~-F.~Chang, J.~N.~Ng and J.~M.~S.~Wu,
  Phys.\ Lett.\ B {\bf 730}, 347 (2014)
  [arXiv:1310.6513 [hep-ph]].


\bibitem{Peccei}
Y.~Chikashige, R.~N.~Mohapatra and R.~D.~Peccei,
Phys. Lett. B {\bf 98}, 265 (1981);
J.~Schechter and J.~W.~F.~Valle,
  Phys.\ Rev.\ D {\bf 25}, 774 (1982).

\bibitem{seesaw}
P.~Minkowsky, Phys. Lett. B {\bf 67}, 421 (1977);
T.~Yanagida {\it Workshop on Unified Theories} (KEK rept)79-18, 95 (1979);
M.~Gell-Mann, P.~Ramond, and R.~Slansky {\it Supergravity} eds. P.~van~Niewenhuizen and D.~Freedman 315 (North Holland, Amsterdam, 1979);
S.~L.~Glashow {\it Cargese Summer Institute on Quarks and Leptons} ed. M.~Levy, p.687 (Plenum Press, N.Y. 1980);
R.~N.~Mohapatra and G.~Senjanovic, Phys. Rev. Lett. {\bf 44} 912 (1980);
 J.~Schechter and J.~W.~F.~Valle,
  Phys.\ Rev.\ D {\bf 22}, 2227 (1980).

\bibitem{snu}
A.~Melchiorri, O.~Mena, S.~Palomares-Ruiz, S.~Pastori, A.~Slosar and M.~Sorel, JCAP {\bf{0901}}, 036 (2009)
[arXiv:0810.5133 [hep-ph]].

\bibitem{RHN}
 L.A.~Anchordoqui, H.~Goldberg, and G.~Steigman, Phys.\ Lett. B {\bf 718}, 1162 (2013).

\bibitem{NTY}
K.~Nakayama, F.~Takahasi, and T.T.~Yanagida, Phys. Lett. B {\bf 697}, 275 (2011).


\bibitem{ADM}
M.~Blennow, E.F.~ Martinez, O.~Mena, J.~Redondo,and P.~Serra, JCAP {\bf 0712}, 022 (2012)
[arXiv: 1203.5803 [hep-ph]].

\bibitem{Hasenkamp:2014hma}
 J.~Hasenkamp,
  arXiv:1405.6736 [astro-ph.CO].

\bibitem{CNW07}
W.~-F.~Chang, J.~N.~Ng and J.~M.~S.~Wu,
Phys. Rev. D {\bf 86}, 033003 (2012), [arXiv: 1206.5047[hep-ph]].



\bibitem{GKMRS13}
P.~P.~Giardino, K.~Kannike, I.~Masina, M.~Raidal and A.~Strumia,
arXiv:1303.3570 [hep-ph].

\bibitem{tome}
A.~Djouadi, Phys. Rept. {\bf 457} 1 (2008).

\bibitem{Crivellin:2013ipa}
 A.~Crivellin, M.~Hoferichter and M.~Procura,
  Phys.\ Rev.\ D {\bf 89}, 054021 (2014)
  [arXiv:1312.4951 [hep-ph]].

\bibitem{BKs}
 L.A.~Anchordoqui, P.B.~Denton, H.~Goldberg, T.C.~Paul, L.H.M.~ da Silva, B.J.~Vlcek and T.J.~Weiler, Phys.\ Rev.\ D {\bf 89} 083513 (2014)
 arXiv: 1312.2547 [hep-ph].

\bibitem{LUX}
D.S.~Akerib,{\it {et al.}}[LUX Collaboration], Phys.\ Rev. \ Lett. {\bf 112} 091303 (2014) arXiv: 1310.8214 [hep-ex].

\bibitem{Daylan:2014rsa}
  T.~Daylan, D.~P.~Finkbeiner, D.~Hooper, T.~Linden, S.~K.~N.~Portillo, N.~L.~Rodd and T.~R.~Slatyer,
  arXiv:1402.6703 [astro-ph.HE].

\bibitem{Martin:2014sxa}
  A.~Martin, J.~Shelton and J.~Unwin,
  arXiv:1405.0272 [hep-ph].

\bibitem{gammaray_spectrum}
  L.~Bergstrom, P.~Ullio and J.~H.~Buckley,
  Astropart.\ Phys.\  {\bf 9}, 137 (1998)
  [astro-ph/9712318];
  J.~L.~Feng, K.~T.~Matchev and F.~Wilczek,
  Phys.\ Rev.\ D {\bf 63} (2001) 045024
  [astro-ph/0008115].


\bibitem{HP}
B.~Pratt and F.Wilczek, [arXiv: 0605188 [hep-ph]].

\bibitem{shadow}
W.F.~Chang, J.N.~Ng, and J.M.S.~Wu, Phys. \ Rev. \ D {\bf 75} 115016 (2007).

\bibitem{SZ}
M.J.~Strassler and K.M.~Zurek, Phys.\ Lett. B {\bf 661} 263 (2008).

\bibitem{KNSW}
A.~Kumar, J.N.~Ng, A.~Spray, and P.T.~Winslow, Phys. \ Rev. D {\bf 88} 075012 (2013).

\bibitem{HK}
J.C.~Helo and S.G.~Kovalenko, Phys. \ Rev. D {\bf 89} 073005 (2014)
[arXiv: 1312.2900 [hep-ph]].

\bibitem{CMS4mu}
CMS Collaboration, Phys. \ Lett. B {\bf 726} 564 (2013).

\end{thebibliography}
\end{document}